\documentclass[11pt,a4paper]{article}
\pdfoutput=1
\usepackage{jcappub}
\usepackage{rotating}
\usepackage{array}
\usepackage{amsmath}
\usepackage[normalem]{ulem}
\usepackage{slashed}
\usepackage{booktabs}
\usepackage[pdftex,table]{xcolor}
\usepackage{units}
\usepackage{multirow}
\usepackage{graphbox}

\arxivnumber{DESY 18-051}

\title{Self-interacting dark matter with a stable vector mediator}

\author{Michael Duerr,}
\author{Kai Schmidt-Hoberg,}
\author{Sebastian Wild}

\affiliation{Deutsches Elektronen-Synchrotron (DESY), Notkestrasse 85, D-22607 Hamburg, Germany}

\emailAdd{michael.duerr@desy.de}
\emailAdd{kai.schmidt-hoberg@desy.de}
\emailAdd{sebastian.wild@desy.de}

\abstract{
Light vector mediators can naturally induce velocity-dependent dark matter self-interactions while at the same time allowing for the correct dark matter relic abundance via thermal freeze-out. If these mediators subsequently decay into Standard Model states such as electrons or photons however, this is robustly excluded by constraints from the Cosmic Microwave Background. We study to what extent this conclusion can be circumvented if the vector mediator is stable and hence contributes to the dark matter density while annihilating into lighter degrees of freedom. We find viable parts of parameter space which lead to the desired self-interaction cross section of dark matter to address the small-scale problems of the collisionless cold dark matter paradigm while being compatible with bounds from the Cosmic Microwave Background and Big Bang Nucleosynthesis observations.
}


\begin{document}
\maketitle
\flushbottom

%%%%%%%%%%%%%%%%%%%%%%%%%%%%%%%%%%%%%%%%%%%%%%%%%%%%%%%%%%%%%%%%%%%%%%%%%%%%%%%
\section{Introduction}
\label{sec:Introduction}
%%%%%%%%%%%%%%%%%%%%%%%%%%%%%%%%%%%%%%%%%%%%%%%%%%%%%%%%%%%%%%%%%%%%%%%%%%%%%%%

Decades of experimental efforts aiming at a discovery of dark matter~(DM) via its non-gravi\-ta\-tion\-al interactions with particles of the Standard Model~(SM) have led to stringent constraints on such couplings, in particular for the popular class of weakly interacting massive particles~(WIMPs)~\cite{Aprile:2017iyp,Ackermann:2015zua,Sirunyan:2017hci}. In contrast, DM self-interactions are largely unconstrained, potentially leading to significant changes in the astrophysical behaviour of DM~\cite{Spergel:1999mh}. 
In fact large DM self-interactions may even be desirable to address a number of discrepancies found in comparing $N$-body simulations of collisionless cold DM with astrophysical observations at small scales (for a recent review see~\cite{Tulin:2017ara}).
In light of this, scenarios in which the DM dominantly couples to particles belonging to a \emph{dark sector} have gained significant attention over the last years (see e.g.~\cite{Feng:2008mu,Foot:2014uba,Berlin:2016gtr,Evans:2017kti}). Interestingly, even a fully decoupled dark sector can lead to falsifiable predictions, e.g.\ to a change in the primordial abundances of elements produced during Big Bang Nucleosynthesis~(BBN)~\cite{Scherrer:1987rr,Hufnagel:2017dgo} or to changes in the Cosmic Microwave Background~(CMB)~\cite{Poulin:2016nat,Bringmann:2018jpr}. 

While large DM self-interactions at small relative velocities are required to address the small-scale problems, there exist rather strong constraints on the DM self-scattering cross section in high-velocity systems such as galaxy clusters~\cite{Markevitch:2003at,Randall:2007ph,Peter:2012jh,Rocha:2012jg,Kahlhoefer:2013dca,Harvey:2015hha,Kaplinghat:2015aga}. A scattering cross section which increases towards smaller velocities is therefore preferred observationally. This behaviour is naturally achieved if a light scalar or vector particle mediates this interaction~\cite{Ackerman:mha,Feng:2009mn,Buckley:2009in,Feng:2009hw,Loeb:2010gj,Aarssen:2012fx,Tulin:2013teo,Kaplinghat:2015aga}. At the same time the DM relic abundance can naturally be set via thermal freeze-out of DM into these mediators.

However, in their simplest forms, these light mediator scenarios are under strong pressure from observations: a vector mediator $Z_\text{D}$ leads to $s$-wave annihilation and if it predominantly decays into SM states such as electrons or photons, the energy injection from late-time annihilations $\psi \bar \psi \rightarrow Z_\text{D} Z_\text{D} \rightarrow \text{SM}$ generically violates the stringent bounds obtained from the CMB~\cite{Bringmann:2016din,Cirelli:2016rnw}. For a scalar mediator, on the other hand, the annihilation is $p$-wave suppressed such that bounds from the CMB are avoided. Nevertheless, strong bounds from direct detection experiments on the coupling to SM states imply late decays of the scalar, which in turn can spoil the successful predictions of standard BBN~\cite{Kaplinghat:2013yxa,Kainulainen:2015sva,BBNfuture}.

A number of possibilities to circumvent these bounds have been discussed for both the vector and scalar cases. To avoid constraints for the vector mediator one possibility is to have decays into light hidden sector states such as sterile neutrinos, which do not lead to reionisation. In such a setup where DM is converted to {\it dark} radiation, bounds from both BBN~\cite{Hufnagel:2017dgo} as well as the CMB~\cite{Bringmann:2018jpr} can be avoided. Another option would be to have asymmetric DM~\cite{Baldes:2017gzu} or to avoid thermalisation of the visible and hidden sectors, in which case freeze-in production~\cite{Bernal:2015ova} can set the relic abundance and constraints can be circumvented. Suppressing the scattering cross section relevant for direct detection allows to have viable models also for scalar mediators~\cite{Blennow:2016gde,Kahlhoefer:2017umn}. 

In this work we study the possibility that the vector mediator $Z_\text{D}$ is stable, in which case the annihilation process $\psi \bar \psi \rightarrow Z_\text{D} Z_\text{D}$ obviously does not lead to energy injection during recombination.
The stability can be achieved either by simply postulating that the kinetic mixing of $Z_\text{D}$ with the SM gauge fields is highly suppressed, or in fact by demanding a dark charge conjugation symmetry~\cite{Ma:2017ucp}. However, in this minimal setup $Z_\text{D}$ freezes out while still being relativistic and, being stable, would overclose the Universe.  

Recently, it has been pointed out~\cite{Ma:2017ucp} that the abundance of a stable vector mediator $Z_\text{D}$ could be sufficiently reduced via annihilations into a lighter state long after the freeze-out of $\psi$. In fact, there is a natural motivation to introduce one more particle in the dark sector: if $Z_\text{D}$ obtains its mass from the breaking of a local $U(1)$ symmetry, the theory contains a \emph{dark Higgs boson} $h_\text{D}$, which (at least at tree-level) has a mass similar to the corresponding gauge boson. For $m_{h_\text{D}} < m_{Z_\text{D}}$, the annihilation $Z_\text{D} Z_\text{D} \rightarrow h_\text{D} h_\text{D}$ can then suppress the late-time $Z_\text{D}$ abundance, and for non-zero mixing between the SM and the dark Higgs boson the latter may decay before dominating the energy density of the Universe.

By construction, the CMB constraints arising from $\psi \bar \psi \rightarrow Z_\text{D} Z_\text{D}$ are avoided; furthermore, the coupling structure of the theory does not permit the annihilation of $\psi \bar \psi$ into a pair of (unstable) dark Higgs bosons $h_\text{D}$ at tree-level. However, the presence of the annihilation channel $\psi \bar \psi \rightarrow Z_\text{D} h_\text{D}$ with the subsequent decay of $h_\text{D}$ still leads to the injection of SM energy into the CMB, and depending on the values of the different couplings involved, this potentially reintroduces the corresponding constraints. Furthermore, the late-time annihilation of the subdominant DM component $Z_\text{D}$ into a pair of dark Higgs bosons can also leave its imprint on the CMB, which is actually well-known to be highly sensitive to even very small annihilation cross sections for DM particles with masses in the MeV range~\cite{Slatyer:2015jla}.

In light of these considerations, we perform a detailed and comprehensive study of the phenomenological viability of this scenario, i.e.~a weak-scale DM particle $\psi$ coupled to a stable vector mediator $Z_\text{D}$, which itself acts as a subdominant DM component. After describing the model in section~\ref{sec:Model}, we discuss the relevant annihilation channels of the two DM species and the corresponding calculation of thermal freeze-out in section~\ref{sec:DMAnnihilation}. In particular, we point out the importance of the conversion processes between the two DM species $\psi$ and $Z_\text{D}$, influencing their cosmological abundances. In section~\ref{sec:constraints}, we first discuss bounds from CMB spectral distortions and BBN on the late-time decay of the dark Higgs boson $h_\text{D}$, before examining the impact of the energy injection during recombination induced by the annihilations of $\psi$ and $Z_\text{D}$. We present our results in section~\ref{sec:Results}, where we pay special attention to the question whether it is possible to have sufficiently strong self-interactions of DM to resolve the small-scale problems mentioned previously, while being consistent with all constraints from the CMB and BBN. Finally, we conclude in section~\ref{sec:conclusions}. Additional material can be found in appendices~\ref{app:FullLagrangian} and~\ref{app:relicdensity}.

%%%%%%%%%%%%%%%%%%%%%%%%%%%%%%%%%%%%%%%%%%%%%%%%%%%%%%%%%%%%%%%%%%%%%%%%%%%%%%%
\section{A simple model}
\label{sec:Model}
%%%%%%%%%%%%%%%%%%%%%%%%%%%%%%%%%%%%%%%%%%%%%%%%%%%%%%%%%%%%%%%%%%%%%%%%%%%%%%%

We extend the SM gauge group by a `dark' gauge symmetry $U(1)_\text{D}$, and introduce a vector-like Dirac fermion $\psi$ as well as a complex scalar $\sigma$ charged under this new symmetry. These dark sector particles are singlets under the SM gauge group, and all SM fields are assumed to transform trivially under $U(1)_\text{D}$. The dark gauge symmetry is then spontaneously broken by a vacuum expectation value (vev) of $\sigma$, resulting in a massive dark gauge boson $Z_\text{D}$ as well as a real scalar $h_\text{D}$.

More precisely, prior to symmetry breaking of the SM and dark gauge group, the Lagrangian of the model is given by
\begin{equation}
 \mathcal{L} = \mathcal{L}_{\widetilde{\text{SM}}} + \mathcal{L}_\text{D} \left( \psi, Z_\text{D}^\mu, \sigma \right) - V \left( \sigma, \Phi \right) \,,
 \label{eq:L}
\end{equation}
with $\mathcal{L}_{\widetilde{\text{SM}}}$ denoting the SM Lagrangian excluding the Higgs potential. The term containing the fermion and gauge boson interactions is given by
\begin{equation}
 \mathcal{L}_\text{D} \left( \psi, Z_\text{D}^\mu, \sigma \right) = i \bar{\psi} \gamma_\mu D^\mu \psi - m_\psi \bar{\psi} \psi + \left( D^\mu \sigma \right)^\ast \left( D_\mu \sigma \right) - \frac{1}{4} F^{\mu\nu}_\text{D} F_{\mu\nu}^\text{D} ,
 \label{eq:LD}
\end{equation}
where
\begin{align}
  D^\mu \psi      &= \left(\partial^\mu - i g_\psi Z^\mu_\text{D} \right) \psi, \\
  D^\mu \sigma &= \left(\partial^\mu - i g_\text{D} Z^\mu_\text{D} \right) \sigma, \\
  F^{\mu\nu}_\text{D} &= \partial^\mu Z_\text{D}^\nu - \partial^\nu Z_\text{D}^\mu.
\end{align}
The $U(1)_\text{D}$ charges (times the gauge coupling) $g_\psi$ and $g_\text{D}$ of the fields $\psi$ and $\sigma$ will be treated as independent parameters of the model. Notice that the mass term of the vector-like fermion $\psi$ is gauge invariant,  and is thus already present prior to symmetry breaking.

Crucially, we have not included a kinetic mixing term $\propto F_\text{D}^{\mu \nu} B_{\mu \nu}$ in eq.~\eqref{eq:LD}, where $B_{\mu \nu}$ denotes the SM hypercharge field strength tensor. After the breaking of $U(1)_\text{D}$ (see below), the presence of this term would allow the massive gauge boson $Z_\text{D}$ to decay into SM states such as $e^\pm$ pairs or photons; as already mentioned in the introduction and explained in more detail in section~\ref{sec:CMB}, basically all of the parameter space of the model leading to significant self-interactions of DM would then be excluded due to constraints on energy injection from DM annihilations during recombination. As pointed out recently in~\cite{Ma:2017ucp}, DM self-interactions might still be viable in such a scenario if the light mediator is stable. From a purely phenomenological point of view, one can thus simply postulate that the dimensionless coupling parameter controlling the kinetic mixing is sufficiently small. For the range of masses of $Z_\text{D}$ considered in this work, a kinetic mixing of the order $\kappa \simeq 10^{-20}$ is necessary to achieve a lifetime equal to the age of the Universe, with stringent bounds from the CMB requiring even smaller values of $\kappa$~\cite{Poulin:2016anj}. Notice that the choice of the kinetic mixing being exactly zero is actually stable under quantum corrections: there are no fermions in the model which are charged both under $U(1)_\text{D}$ as well as under a SM gauge symmetry, and hence all loop-induced contributions to the mixing of $Z_\text{D}$ with the SM gauge bosons vanish.

Alternatively, as pointed out recently in~\cite{Ma:2017ucp}, the kinetic mixing term can be forbidden by imposing  a \emph{dark charge conjugation symmetry}, rendering $Z_\text{D}$ absolutely stable (as long as $m_{Z_\text{D}} < 2 m_\psi$). In the same way as there is the familiar charge conjugation operator $\mathcal{C}$ associated with the SM $U(1)_\text{em}$ group, the dark charge conjugation operator $\mathcal{C}_\text{D}$ changes the signs of the $U(1)_\text{D}$ charges $g_\psi$ and $g_\text{D}$, and furthermore replaces $\sigma$ by $\sigma^\ast$, $Z_\text{D}^\mu$ by $-Z_\text{D}^\mu$  as well as $\psi$ by the charge-conjugated spinor $\psi^\text{C}$. If, in contrast to $\mathcal{C}$, nature is symmetric with respect to dark charge conjugation, the kinetic mixing operator $F_\text{D}^{\mu \nu} B_{\mu \nu}$ is forbidden. Notice that this symmetry is still present after the spontaneous breaking of $U(1)_\text{D}$ via a vev of $\sigma$. 

Finally, in the Lagrangian given by eq.~\eqref{eq:L}, $V \left( \sigma, \Phi \right)$ denotes  the most general scalar potential involving the SM singlet $\sigma$ and the SM Higgs doublet $\Phi$:
\begin{equation}
 V \left( \sigma, H \right) = -\mu_\text{D}^2 \sigma^\ast \sigma + \frac{1}{2} \lambda_\text{D} \left( \sigma^\ast \sigma \right)^2 - \mu_h^2 \Phi^\dagger \Phi + \frac12 \lambda_h \left( \Phi^\dagger \Phi \right)^2+ \lambda_{h\text{D}} \left( \sigma^\ast \sigma \right) \left( \Phi^\dagger \Phi \right) \,.
 \label{eq:scalarpotential_beforeSB}
\end{equation}
After spontaneous breaking of the electroweak and dark gauge symmetry, the scalar fields can be parametrised in unitary gauge as
\begin{equation}
 \sigma = (v_\text{D} + H_\text{D})/\sqrt{2} \text{ and } \Phi = (0, (v_h + H)/\sqrt{2})^T.
 \label{eq:scalars_afterSB}
\end{equation}
In the following, we eliminate $\lambda_\text{D}$ and $\lambda_h$ from the scalar potential~\eqref{eq:scalarpotential_beforeSB} by using $v_h \simeq \unit[246]{GeV}$ and treating the dark Higgs vev $v_\text{D}$ as a free parameter. For a given choice of the gauge coupling $g_\text{D}$, the latter is in one-to-one correspondence with the gauge boson mass $m_{Z_\text{D}} = g_\text{D} v_\text{D}$. 

The presence of the portal term proportional to $\lambda_{h \text{D}}$ in the scalar potential leads to a mixing of $H_\text{D}$ and $H$; we denote the corresponding mass eigenstates by $h_\text{D}$ and $h$. Assuming $\lambda_{h\text{D}} \ll 1$, $m_{h_\text{D}} \ll m_h$, the mixing angle is given by  $\theta \simeq \lambda_{h\text{D}} v_\text{D} v_h/m_h^2$, where $m_h \simeq \unit[125]{GeV}$ is the mass of the SM Higgs boson $h$. While in the absence of the kinetic mixing term $Z_\text{D}$ is stable, the dark Higgs boson $h_\text{D}$ can decay into SM particles with a rate proportional to $\theta^2$. Further details, in particular the full Lagrangian including the scalar potential  after symmetry breaking can be found in appendix~\ref{app:FullLagrangian}.

For the purpose of our phenomenological analysis, a point in the parameter space of the model after symmetry breaking is then fully specified by the free parameters
\begin{equation}
 m_{Z_\text{D}}, m_\psi, m_{h_\text{D}}, g_\text{D}, g_\psi, \lambda_{h\text{D}} \,.
\end{equation}
Note that as long as the dimensionless couplings $g_\text{D}$ and $\lambda_\text{D}$ are of order one, $m_{Z_\text{D}}$ and $m_{h_\text{D}}$ are expected to be of the same order of magnitude. On the other hand, the tree-level mass of $\psi$ is not related to the breaking of $U(1)_\text{D}$, and thus can be naturally at a different scale.

%%%%%%%%%%%%%%%%%%%%%%%%%%%%%%%%%%%%%%%%%%%%%%%%%%%%%%%%%%%%%%%%%%%%%%%%%%%%%%%
\section{Annihilation channels of dark matter and freeze-out calculation}
\label{sec:DMAnnihilation}
%%%%%%%%%%%%%%%%%%%%%%%%%%%%%%%%%%%%%%%%%%%%%%%%%%%%%%%%%%%%%%%%%%%%%%%%%%%%%%%

The scenario  introduced in the previous section involves two stable neutral particles which contribute to the observed density of DM: the Dirac fermion $\psi$ as well as the massive gauge boson $Z_\text{D}$. In the following, we discuss the main qualitative aspects of the freeze-out process of these DM particles; additional technical details of our numerical implementation can be found in appendix~\ref{app:relicdensity}.

\begin{figure}[tb]
\centering
\includegraphics[align=c,scale=.5]{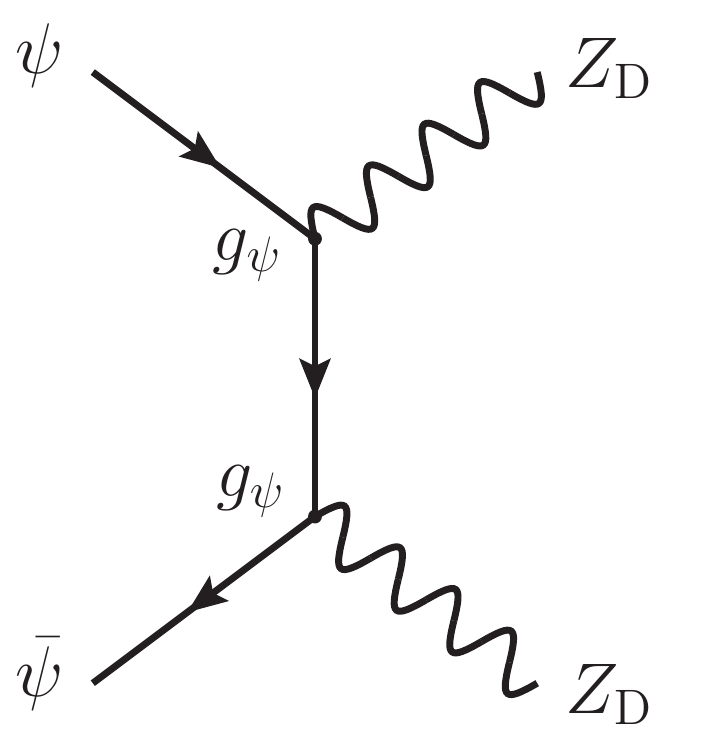}\hspace{6ex}
\includegraphics[align=c,scale=.5]{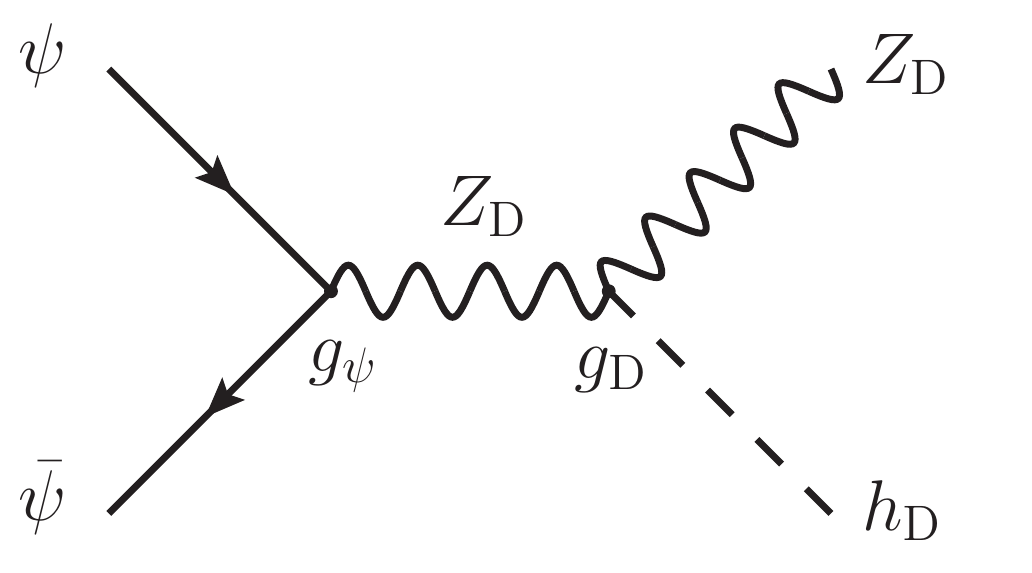}
\caption{Feynman diagrams visualising the annihilation channels %for the Dirac fermion $\psi$: 
$\psi \bar{\psi} \to Z_\text{D} Z_\text{D}$ (left) and $ \psi \bar{\psi} \to Z_\text{D} h_\text{D}$ (right). The corresponding cross sections are given in Eqs.~\eqref{eq:psipsiZDZD} and \eqref{eq:psipsiZDhD}.}
\label{fig:psi_annihilation}
\end{figure}

\begin{figure}[tb]
\centering
\includegraphics[align=c,scale=.5]{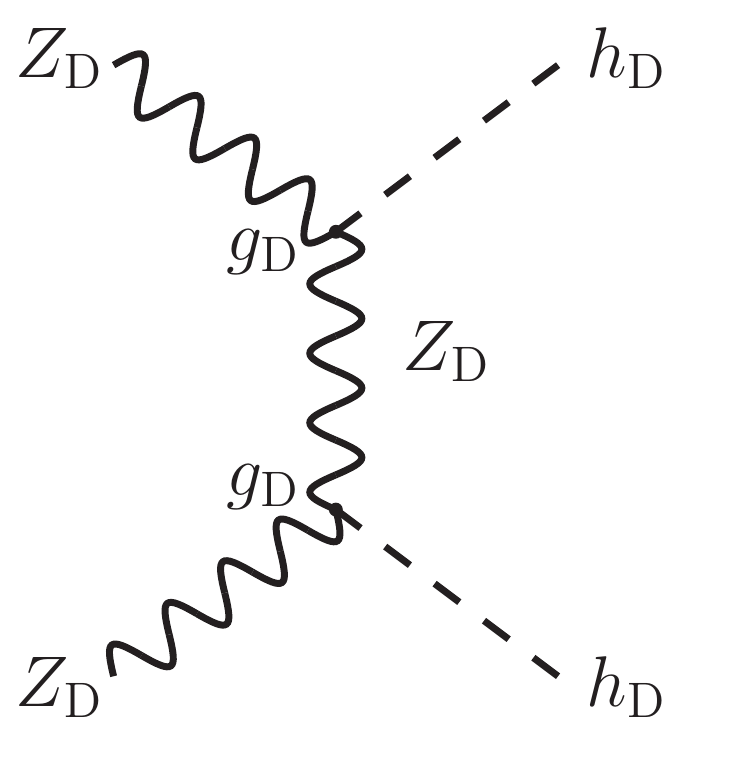}\hspace{6ex}
\includegraphics[align=c,scale=.5]{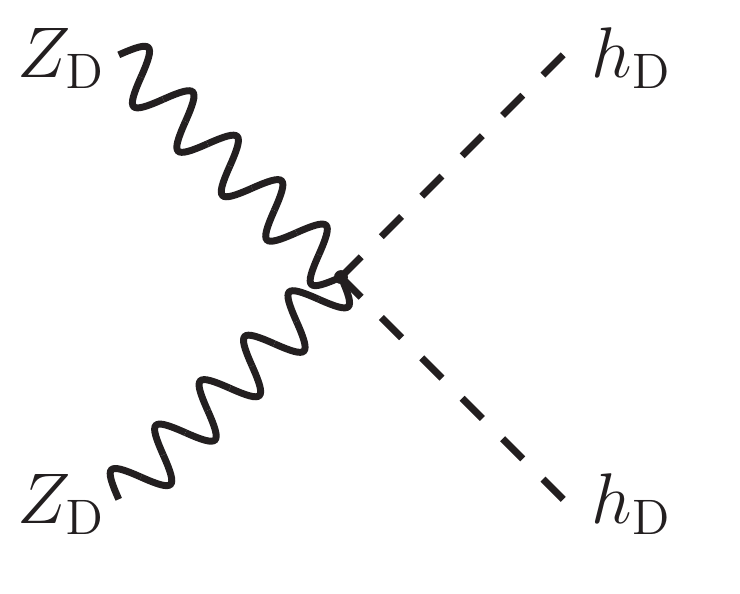}\hspace{6ex}
\includegraphics[align=c,scale=.5]{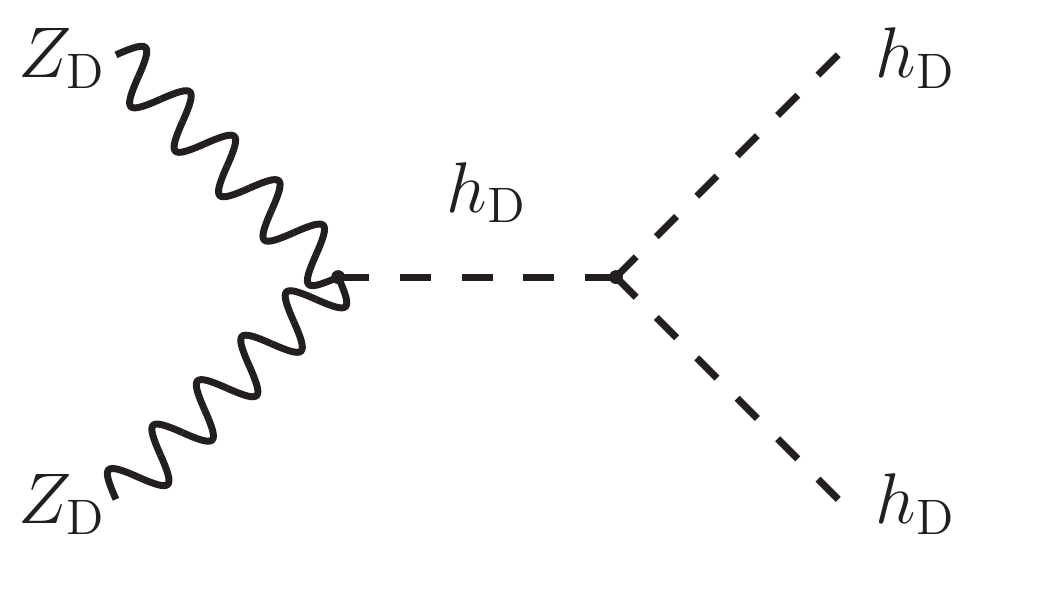}
\caption{Feynman diagrams depicting the annihilation of the massive gauge boson $Z_\text{D}$. The cross section for the process $ Z_\text{D} Z_\text{D} \to h_\text{D} h_\text{D}$ is given in Eq.~\eqref{eq:ZDZDhdhD}.}
\label{fig:ZD_annihilation}
\end{figure}

We focus our analysis on regions in parameter space where $m_{Z_\text{D}} \ll m_\psi$: this is a necessary condition for obtaining a self-interaction cross section of $\psi$ which is large enough to lead to interesting astrophysical signatures. 
The heavy DM particle $\psi$ can then self-annihilate via two possible channels (see Fig.~\ref{fig:psi_annihilation} for the corresponding Feynman diagrams):
\begin{align}
 \psi \bar{\psi} \to Z_\text{D} Z_\text{D} \quad \text{with } (\sigma v)_{\psi \bar{\psi} \to Z_\text{D} Z_\text{D}}^\text{tree} \simeq \frac{g_\psi^4}{16 \pi m_\psi^2} \,, \label{eq:psipsiZDZD}\\
\psi \bar{\psi} \to Z_\text{D} h_\text{D} \quad \text{with } (\sigma v)_{\psi \bar{\psi} \to Z_\text{D} h_\text{D}}^\text{tree} \simeq \frac{g_\text{D}^2 g_\psi^2}{64 \pi m_\psi^2} \,, \label{eq:psipsiZDhD}
\end{align}
where the tree-level expressions $(\sigma v)^\text{tree}$ for the annihilation cross sections are given in the limit $m_{Z_\text{D}} \ll m_\psi$ and $v \ll 1$. Note that the latter process leads to significant constraints from the CMB via the decay of the dark Higgs into SM states (see Section~\ref{sec:CMB}), which have not been considered in~\cite{Ma:2017ucp}. The annihilation of $\psi \bar \psi$ into a pair of dark Higgs bosons, on the other hand, is strongly suppressed as it only proceeds via a one-loop diagram and furthermore vanishes in the $s$-wave limit $v \to 0$. In our numerical calculation, we take into account Sommerfeld enhancement in the annihilation processes~(\ref{eq:psipsiZDZD}) and~(\ref{eq:psipsiZDhD}), arising from the multiple exchange of $Z_\text{D}$ bosons in the $\psi \bar \psi$ initial state (see appendix~\ref{app:relicdensity} for details). Moreover, for $m_{h_\text{D}} < m_{Z_\text{D}}$ the massive gauge boson $Z_\text{D}$ can annihilate via 
\begin{align}
 Z_\text{D} Z_\text{D} \to h_\text{D} h_\text{D} \quad \text{with } (\sigma v)_{ Z_\text{D} Z_\text{D} \to h_\text{D} h_\text{D}} \simeq \frac{g_\text{D}^4 \sqrt{1-r} \left( 44 - 20 r + 9 r^2 - 8 r^3 + 2 r^4 \right)}{9 \pi m_{Z_\text{D}}^2 \left( 8 - 6 r + r^2 \right)^2 } \,,
 \label{eq:ZDZDhdhD}
\end{align}
where $r = m_{h_\text{D}}^2 / m_{Z_\text{D}}^2$.\footnote{This expression differs from the one given in Ref.~\cite{Ma:2017ucp}.} The corresponding Feynman diagrams are shown in Fig.~\ref{fig:ZD_annihilation}.

At large temperatures, these annihilation processes lead to chemical equilibrium between the dark sector particles $\psi$, $Z_\text{D}$ and $h_\text{D}$. Furthermore, in the following we assume the portal coupling $\lambda_{h \text{D}}$ to be sufficiently large such that the initial temperatures of the dark and visible sectors are identical; the precise choice for $\lambda_{h \text{D}}$ will be discussed in more detail in section~\ref{sec:BBN}. The cosmological evolution of the DM particles $\psi$ and $Z_\text{D}$ down to smaller temperatures is then described by a set of two coupled Boltzmann equations for the number densities $n_\psi$ and $n_{Z_\text{D}}$. As described in more detail in appendix~\ref{app:relicdensity}, we compute the present-day  abundances $\Omega_\psi h^2$~(defined to be the sum of the abundances of $\psi$ and $\bar \psi$) and $\Omega_{Z_\text{D}} h^2$ by solving these equations numerically using a modified version of \texttt{MicrOMEGAs v4.3.5}~\cite{Belanger:2006is,Belanger:2014vza}, additionally taking into account the Sommerfeld enhancement as well as thermal decoupling of the dark and visible sector at a temperature $T_\text{dec}$. 

Qualitatively, the freeze-out process can be understood as follows: at $T \simeq m_\psi/25$, the annihilation processes given in eqs.~\eqref{eq:psipsiZDZD} and~\eqref{eq:psipsiZDhD} stop being efficient, and the heavy DM particle $\psi$ freezes out, i.e.~$n_\psi/s$ becomes constant. However, the lighter DM particle $Z_\text{D}$ remains in chemical equilibrium with the dark Higgs boson down to much smaller temperatures $T \simeq m_{Z_\text{D}} / x_{f}$, with $x_f \simeq 15-50$. The precise value of $x_f$, and thus the final abundance of $Z_\text{D}$ depends on the strength of various annihilation channels: besides the usual self-annihilation $Z_\text{D} Z_\text{D} \to h_\text{D} h_\text{D}$, also processes involving the already frozen-out DM particle $\psi$ have to be taken into account, leading to additional terms in the Boltzmann equation for $n_{Z_\text{D}}$. Concretely, these are the conversion process $\psi Z_\text{D} \to \psi h_\text{D}$ as well as the annihilation channels $\psi \bar \psi \to Z_\text{D} Z_\text{D}$ and $\psi \bar \psi \to Z_\text{D} h_\text{D}$. Notice that even though during the freeze-out of $Z_\text{D}$ the latter processes are already too weak in order to keep $\psi$ in equilibrium, they nevertheless can be important for the evolution of $n_{Z_\text{D}}$. A more detailed discussion of this point can be found in appendix~\ref{app:relicdensity}.

%%%%%%%%%%%%%%%%%%%%%%%%%%%%%%%%%%%%%%%%%%%%%%%%%%%%%%%%%%%%%%%%%%%%%%%%%%%%%%%
\section{Observational constraints}
\label{sec:constraints}
%%%%%%%%%%%%%%%%%%%%%%%%%%%%%%%%%%%%%%%%%%%%%%%%%%%%%%%%%%%%%%%%%%%%%%%%%%%%%%%

\subsection[Bounds on the decay of \texorpdfstring{$h_\text{D}$}{hD} from CMB spectral distortions and BBN]{Bounds on the decay of \texorpdfstring{$\boldsymbol{h}_\text{D}$}{hD} from CMB spectral distortions and BBN}
\label{sec:BBN}

Being in thermal equilibrium with the SM heat bath at early times, the dark Higgs boson $h_\text{D}$ generically has a significant abundance prior to its decay. As we are interested in a scenario with $m_{h_\text{D}} < m_{Z_\text{D}} \lesssim \unit[100]{MeV}$, it decays either dominantly into $e^+ e^-$ (for $m_{h_\text{D}} > 2 m_e$) or into $\gamma \gamma$ (for $m_{h_\text{D}} < 2 m_e$), with a lifetime $\tau_\phi$ taken from~\cite{Bezrukov:2009yw,Alekhin:2015byh}. If these decay products are injected at redshifts $z \lesssim 2 \times 10^6$, they do not fully thermalise with the background photons, and thus lead to spectral distortions in the CMB~\cite{Zeldovich:1969ff,Hu:1993gc,Chluba:2011hw}. In the context of our scenario, this excludes all regions of parameter space with $\tau_{h_\text{D}} \gtrsim \unit[10^5]{s}$~\cite{Poulin:2016anj}.\footnote{This bound can be circumvented if the dark Higgs is stable on cosmological timescales and sufficiently light such that it does not contribute significantly to the present-day density of DM. In fact all CMB bounds from late time energy injection will be evaded in this case. In the following we do not further consider this part of the parameter space, and focus on the case where $m_{h_\text{D}}$ and $m_{Z_\text{D}}$ are of similar order of magnitude.} Even for a scalar portal coupling $\lambda_{h\text{D}}$ of order one, this bound is generically violated if the dark Higgs has a mass below $2 m_e$ and thus can only decay into a pair of photons at one loop. As we still want to keep $m_{h_\text{D}} < m_{Z_\text{D}}$ in order to allow for the annihilation process $Z_\text{D} Z_\text{D} \to h_\text{D} h_\text{D}$ to deplete the abundance of $Z_\text{D}$, in the following we will fix $m_{h_\text{D}} = \unit[1.5]{MeV} > 2 m_e$, and only consider vector boson masses $m_{Z_\text{D}} \gtrsim \unit[2]{MeV}$. Notice that as long as $m_{Z_\text{D}} \gtrsim m_{h_\text{D}}$, the precise value of the dark Higgs boson mass does not impact the phenomenology elsewhere, in particular neither the CMB constraints on energy injection from DM annihilation during recombination nor the self-interaction cross section of $\psi$.  When presenting our results in section~\ref{sec:Results}, we will indicate in which regions of parameter space the lifetime of $h_\text{D}$ for this choice of $m_{h_\text{D}}$ nevertheless exceeds $\unit[10^5]{s}$, and is thus excluded by the constraints on CMB spectral distortions.

The decay of $h_\text{D}$ in the early Universe is also constrained by the excellent agreement of the observed primordial abundances of light elements with the predictions from BBN. In general, BBN can be affected by additional stable or decaying particles present at temperatures $T \lesssim \unit[10]{MeV}$~\cite{Cyburt:2015mya,Patrignani:2016xqp}. More specifically, the scenario discussed in this work potentially modifies the primordial nuclear abundances in two ways:
\begin{itemize}
\item[(i)] If the dark Higgs $h_\text{D}$ decays well after BBN, its electromagnetic decay products can photo-disintegrate nuclei, in particular deuterium and helium.
\item[(ii)] If $Z_\text{D}$ and/or $h_\text{D}$ are still in thermal equilibrium at $T \simeq \unit[10]{MeV}$, they provide a contribution to $\Delta N_\text{eff}$ and thus enhance the expansion rate during BBN.
\end{itemize}
The first bound potentially constrains regions of parameter space where $\tau_{h_\text{D}} \gtrsim \unit[10^4]{s}$; for smaller lifetimes, the cascade of the electromagnetic decay products caused by interactions with CMB photons leads to a cutoff of the corresponding photon spectrum below the photo-disintegration threshold $E_\text{dis} = \unit[2.2]{MeV}$ of deuterium~\cite{Jedamzik:2006xz,Berger:2016vxi}. However, for our choice $m_{h_\text{D}} = \unit[1.5]{MeV}$ as motivated above from the constraints on CMB spectral distortions, the electromagnetic cascade induced by the electrons and positrons produced in the decay of $h_\text{D}$ anyway only lead to photons with energies below $E_\gamma \simeq \unit[0.75]{MeV}$, which are unable to photo-disintegrate nuclei even for lifetimes $\tau_{h_\text{D}} \gg \unit[10^4]{s}$. Consequently, for our choice of $m_{h_\text{D}}$, the BBN bound (i) is automatically avoided.

\begin{figure}[tb]
\centering
\includegraphics[scale=1.0]{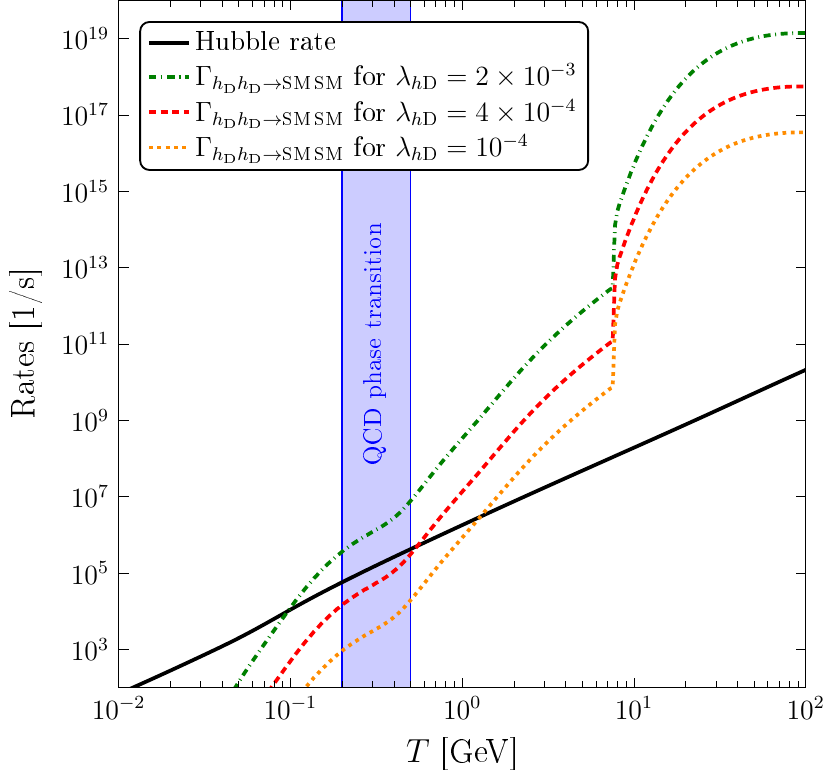}
\caption{Reaction rate $\Gamma_{h_\text{D} h_\text{D} \rightarrow \text{SM SM}}$ for different choices of $\lambda_{h\text{D}}$. As explained in the text, choosing $\lambda_{h\text{D}} \lesssim 4 \times 10^{-4}$ leads to thermal decoupling of the dark and visible sectors prior to the QCD phase transition, and thus to a significantly reduced contribution of the dark sector particles to $\Delta N_\text{eff}$.}
\label{fig:hDhD_rates}
\end{figure}

The constraint (ii) from the increased Hubble rate during BBN depends critically on the temperature of the dark sector $T_\text{D}$ at $T \simeq \unit[10]{MeV}$. The process most relevant for keeping the dark and visible sectors in thermal contact (leading to $T_\text{D} = T$) is the annihilation of the dark Higgs $h_\text{D}$ into SM particles. The corresponding reaction rate $\Gamma_{h_\text{D} h_\text{D} \to \text{SM SM}}(T)$ as a function of temperature is shown in Fig.~\ref{fig:hDhD_rates}  for different choices of the parameter $\lambda_{h\text{D}}$ appearing in the scalar potential~\eqref{eq:scalarpotential_beforeSB}. For $T \gtrsim m_h/2$, the dominant process establishing equilibrium is the production of an on-shell SM Higgs boson in the s-channel which, even for rather small values of $\lambda_{h\text{D}}$, guarantees chemical equilibrium at these temperatures. For smaller $T$, this process gets exponentially suppressed and the annihilation rate $\Gamma_{h_\text{D} h_\text{D} \rightarrow \text{SM SM}}$ rapidly decreases\footnote{For $T \lesssim 5\,$GeV, the light SM quarks are no longer the appropriate degrees of freedom in the thermal bath. Following~\cite{Cline:2013gha}, in this regime the annihilation cross section for $h_\text{D} h_\text{D} \to \text{SM SM}$ at a given center-of-mass energy $\sqrt{s}$ can be expressed in terms of the width of a (hypothetical) scalar particle with mass $m_\star = \sqrt{s}$, which in turn we take from~\cite{Alekhin:2015byh}.}, until eventually the dark and visible sectors decouple at a temperature $T_\text{dec}$, which we define via $\Gamma_{h_\text{D} h_\text{D} \rightarrow \text{SM SM}}(T_\text{dec}) = H(T_\text{dec})$. As can be seen from Fig.~\ref{fig:hDhD_rates}, by choosing $\lambda_{h\text{D}} \lesssim 4 \times 10^{-4}$, this decoupling happens prior to the QCD phase transition, i.e.~$T_\text{dec} \gtrsim \unit[500]{MeV}$. The visible sector is then heated with respect to the dark sector, reducing the relative contribution of the dark sector particles to the energy density. Quantitatively, the impact of $Z_\text{D}$ and $h_\text{D}$ on the Hubble rate during BBN can be parametrised in terms of the equivalent number of additional neutrino species:
\begin{align}
\Delta N_\text{eff} (T \simeq \unit[10]{MeV}) \simeq 4 \cdot \left( \frac{g_\text{SM} (\unit[10]{MeV})}{g_\text{SM} (T_\text{dec})} \right)^{4/3} \lesssim 0.27 \,.
\label{eq:DeltaNeff}
\end{align}
Here we conservatively assumed that both $Z_\text{D}$ and $h_\text{D}$ are relativistic degrees of freedom during BBN; for $m_{Z_\text{D}} \gtrsim \unit[10]{MeV}$ the abundance of $Z_\text{D}$ during BBN is already Boltzmann suppressed, and the contribution to $\Delta N_\text{eff}$ is even smaller. Using the most recent information on the baryon-to-photon ratio inferred from the CMB as well as updated nuclear reaction rates, the upper limit on extra radiation during BBN is found to be $\Delta N_\text{eff} < 0.2 (0.36)$ at $2\sigma (3\sigma)$~\cite{Cyburt:2015mya}. Given the  significant impact of systematic uncertainties on deriving this limit, we conclude that the maximal contribution to $\Delta N_\text{eff}$ predicted by our scenario, as given by eq.~\eqref{eq:DeltaNeff}, might be in (mild) tension with BBN observations, but is certainly not robustly ruled out. A detailed analysis of BBN constraints on MeV-scale particles decaying into SM states, going beyond the simple estimate of $\Delta N_\text{eff}$ via eq.~\eqref{eq:DeltaNeff} will appear elsewhere~\cite{BBNfuture}.

As outlined above, this conclusion holds as long as $\lambda_{h\text{D}} \lesssim 4 \times 10^{-4}$, such that the dark and visible sectors decouple before the QCD phase transition. On the other hand, by choosing $\lambda_{h\text{D}}$ too small, the lifetime of the dark Higgs boson can get larger than $\tau_{h_\text{D}} \gtrsim \unit[10^{5}]{s}$, violating the bound from CMB spectral distortions as discussed at the beginning of this section. In order to weaken this constraint as much as possible, we fix $\lambda_{h\text{D}} = 4 \times 10^{-4}$ in the following, i.e.~we choose the maximal value compatible with the constraint on the Hubble rate during BBN.\footnote{This choice of $\lambda_{h\text{D}}$ leads to an invisible decay width $\Gamma_{h \rightarrow h_\text{D} h_\text{D}} = \lambda_{h\text{D}}^2 v_h^2/(16 \pi m_h) \simeq 3.8 \times 10^{-4} \times \Gamma_h^\text{tot}$ of the SM Higgs, which is well below the constraint from the latest LHC data~\cite{Khachatryan:2016whc}. Furthermore, depending on the vev $v_\text{D}$ of the scalar field $\sigma$, the corresponding mixing angle $\theta$ of the dark Higgs boson can be in the range where it might significantly alter the duration of the neutrino pulse from SN1987a~\cite{Raffelt:1987yu, Krnjaic:2015mbs}. However, in view of the still large systematic uncertainties inherent in deriving the corresponding bounds, we do not consider them in the following discussion; a dedicated analysis of this point would certainly be interesting.} With this value for $\lambda_{h\text{D}}$, the lifetime of the dark Higgs will exceed $\tau_{h_\text{D}} = \unit[10^{5}]{s}$ in some parts of the parameter regions considered in the numerical analysis in Sec.~\ref{sec:Results}. We indicate the corresponding regions in all plots, but note that they are independently excluded by other constraints.

\subsection{CMB constraints on energy injection during recombination}
\label{sec:CMB}

The prime motivation for postulating the stability of $Z_\text{D}$ has been to avoid the constraints arising from energy injection during recombination due to the annihilation process $\psi \bar \psi \to Z_\text{D} Z_\text{D}$. However, in our scenario the heavy DM particle can also annihilate via $\psi \bar{\psi} \rightarrow Z_\text{D} h_\text{D}$, potentially reintroducing the CMB constraints due to the subsequent decays of the dark Higgs boson into SM states. Moreover, also late-time annihilations  $Z_\text{D} Z_\text{D} \rightarrow h_\text{D} h_\text{D}$  lead to energy injection into the CMB, which, depending on the fraction of DM made up of $Z_\text{D}$, might also be in conflict with observations.
 
The annihilation cross section for $\psi \bar{\psi} \rightarrow Z_\text{D} h_\text{D}$ during recombination is given by
\begin{align}
(\sigma v)_{\psi \bar{\psi} \to Z_\text{D} h_\text{D}}^\text{CMB} \equiv S_s(v) \cdot (\sigma v)_{\psi \bar{\psi} \to Z_\text{D} h_\text{D}}^\text{tree},
\label{eq:sigmavCMB_psipsi}
\end{align} 
where $(\sigma v)_{\psi \bar{\psi} \to Z_\text{D} h_\text{D}}^\text{tree}$ is the tree-level cross section given in eq.~\eqref{eq:psipsiZDhD}, and $S_s(v)$  is the $s$-wave Sommerfeld enhancement factor corresponding to the multiple exchange of $Z_\text{D}$ in the initial state, which is provided in eq.~\eqref{eq:Ss}. The relative velocity $v$ during recombination entering eq.~\eqref{eq:sigmavCMB_psipsi} can be conservatively estimated by using an upper bound on the kinetic decoupling temperature of DM from Lyman-$\alpha$ observations~\cite{Croft:1997jf, Croft:2000hs}, resulting in~\cite{Bringmann:2016din}
\begin{equation}
 v \lesssim 2 \times 10^{-7} \left( \frac{m_\psi}{\unit[100]{GeV}}\right)^{-1/2}.
\end{equation}
We have explicitly confirmed that the precise value of $v$ does not affect our results as long as it satisfies this bound, since the Sommerfeld enhancement is already saturated for these velocities. A given point in parameter space is then excluded by CMB data if
\begin{align}\label{eq:sigmav_NNX_bound}
\frac12 \cdot (\sigma v)_{\psi \bar{\psi} \to Z_\text{D} h_\text{D}}^\text{CMB} > (\sigma v)_{4e^\pm}^{\text{(upper bound)}}\left( m_\psi \right) \cdot \left(\frac{\Omega_\text{DM} h^2}{\Omega_{\psi} h^2}\right)^2 \,.
\end{align}
Here the factor $1/2$ on the left hand side accounts for the fact that due to the stability of $Z_\text{D}$ only half of the energy is transferred into electrons and positrons affecting reionisation. Furthermore, $(\sigma v)_{4e^\pm}^{\text{(upper bound)}}(m_\psi)$ is the upper bound on the annihilation cross section of DM into a final state containing two electrons and two positrons, obtained under the assumption that $\psi$ constitutes all of the observed DM. We take this bound as a function of $m_{\psi}$ from~\cite{Slatyer:2015jla}, after multiplying it by a factor of two due to the Dirac nature of $\psi$. 
Finally, the last factor in eq.~\eqref{eq:sigmav_NNX_bound} takes into account the suppression of the bound if $\psi$ does not constitute all of the observed DM, with $\Omega_\text{DM} h^2 \simeq 0.12$ being the total DM abundance~\cite{Ade:2015xua}.

Similarly, the  energy injection during recombination due to annihilations of $Z_\text{D}$ excludes parts of the parameter space where
\begin{align}
(\sigma v)_{Z_\text{D} Z_\text{D} \rightarrow h_\text{D} h_\text{D}} > (\sigma v)_{4e^\pm}^{\text{(upper bound)}} \left( m_{Z_\text{D}} \right) \cdot \left(\frac{\Omega_\text{DM} h^2}{\Omega_{Z_\text{D}} h^2}\right)^2 \,,
\label{eq:sigmav_ZDZDX_bound}
\end{align}
with $(\sigma v)_{Z_\text{D} Z_\text{D} \rightarrow h_\text{D} h_\text{D}}$ given by eq.~\eqref{eq:ZDZDhdhD}. Notice that  in contrast to the self-annihilation of $\psi$, for the values of $m_{Z_\text{D}}/m_{h_\text{D}}$ considered in this work this process is not subject to Sommerfeld enhancement, and the corresponding cross section can simply be evaluated in the limit $v \rightarrow 0$.

\subsection{Self-interactions of dark matter}
\label{sec:selfinteractions}

Via its coupling to the light mediator $Z_\text{D}$, the DM particle $\psi$ can experience significant rates of self-scattering, even for weak couplings $g_\psi \lesssim 1$~\cite{Buckley:2009in,Feng:2009hw}. This process can have important consequences for the distribution of DM in various astrophysical systems: it can transform cuspy profiles of DM halos into cored ones~\cite{Yoshida:2000uw,Dave:2000ar} or more generally lead to a large diversity of DM profiles once baryonic effects are taken into account~\cite{Kamada:2016euw}. It may even lead to spectacular displacement signatures in merging galaxy clusters~\cite{Williams:2011pm,Dawson:2011kf,Kahlhoefer:2013dca} if the scattering cross section is only mildly suppressed at large velocities (see~\cite{Tulin:2017ara} for a recent review on the subject).

For a large class of astrophysical objects, a good proxy for the impact of DM self-interactions is the momentum transfer cross section $\sigma_\text{T}$, defined via~\cite{Kahlhoefer:2013dca,Kahlhoefer:2017umn}
\begin{align}\label{eq:transferCrossSection}
 \sigma_\text{T} &\equiv \frac{1}{2} \left( \sigma_\text{T}^{\psi \psi} + \sigma_\text{T}^{\psi \bar \psi} \right) \, \, ,\text{ with } \nonumber \\
  \sigma_\text{T}^{{\psi \psi},{\psi \bar \psi}} &\equiv 2 \pi \int_{-1}^1 \left(\frac{\text{d}\sigma}{\text{d}\Omega}\right)^{{\psi \psi},{\psi \bar \psi}} \left(1 - \left| \cos\theta \right| \right) \text{d}\cos\theta .
\end{align}
Here, $(\text{d}\sigma/\text{d}\Omega)^{\psi \psi}$ and $(\text{d}\sigma/\text{d}\Omega)^{\psi \bar \psi}$  denote the differential cross sections for elastic scattering of $\psi \psi$ and $\psi \bar \psi$, respectively. We compute those by adapting the procedure outlined in~\cite{Kahlhoefer:2017umn} for DM interacting with a scalar mediator to the case of a vector mediator. In particular, we take into account non-perturbative effects related to multiple exchange of $Z_\text{D}$ by solving the corresponding Schr\"odinger equation for a Yukawa-like scattering potential, properly taking into account the quantum indistinguishability of identical particles participating in the scattering process. For $g_\psi^2 m_\psi/(4 \pi m_{Z_\text{D}}) \ll 1$, the non-perturbative effects are negligible and our results match the analytical expressions given in~\cite{Kahlhoefer:2017umn} for the Born regime (which are identical for scalar and vector mediators). On the other hand, for $m_\psi v/m_{Z_\text{D}} \gtrsim 5$ solving the Schr\"odinger equation becomes not feasible, and we employ the results from~\cite{Cyr-Racine:2015ihg} for the scattering cross section in the classical regime.

Crucially, in the regime where non-perturbative effects are important, the momentum transfer cross section $\sigma_\text{T}$ typically is enhanced for small velocities $v$ of the DM particles. Hence, one naturally expects larger effects of the DM self-scattering process in systems with small velocity dispersions such as dwarf galaxies (where $v \simeq 30\,\text{km}\,\text{s}^{-1}$), and thus it is easier to satisfy the upper bounds on $\sigma_\text{T}$ from observations of galaxy clusters (where $v \simeq 1000\,\text{km}\,\text{s}^{-1}$). However, both the cross section required in order to transform cusps in dwarf galaxies into cored profiles~\cite{Vogelsberger:2012ku,Rocha:2012jg,Zavala:2012us,Tulin:2013teo,Kaplinghat:2015aga}, as well as the largest value of $\sigma_\text{T}/m_\psi$ compatible with constraints from merging galaxy clusters~\cite{Randall:2007ph,Kahlhoefer:2015vua,Harvey:2015hha,Wittman:2017gxn} are still under debate. In light of this, and in order to bracket all of the potentially interesting range of momentum transfer cross sections at small scales, in section~\ref{sec:Results} we will show which regions in parameter space lead to $0.1\,\text{cm}^2/\text{g} < \sigma_\text{T}/m_\psi < 10 \, \text{cm}^2/\text{g}$ at $v \simeq 30\,\text{km}\,\text{s}^{-1}$, and use the rather conservative upper bound $\sigma_\text{T}/m_\psi < 1  \,\text{cm}^2/\text{g}$ at the scale of galaxy clusters, $v \simeq 1000\,\text{km}\,\text{s}^{-1}$.\footnote{Both the preferred range for $\sigma_\text{T}/m_\psi$ at small scales as well as the upper bound at scales of galaxy clusters have been derived assuming that all of the observed DM is self-interacting, while in our scenario $Z_\text{D}$ does not experience significant self-interactions. However, as we will see in section~\ref{sec:Results}, in all regions of the parameter space where the self-interaction cross section of $\psi$ is within the range of interest, one has $\Omega_{Z_\text{D}} h^2 \ll \Omega_\psi h^2 \simeq 0.12$, and hence the astrophysical behaviour of DM is dominated by the properties of $\psi$ alone.}

%%%%%%%%%%%%%%%%%%%%%%%%%%%%%%%%%%%%%%%%%%%%%%%%%%%%%%%%%%%%%%%%%%%%%%%%%%%%%%%
\section{Results}
\label{sec:Results}
%%%%%%%%%%%%%%%%%%%%%%%%%%%%%%%%%%%%%%%%%%%%%%%%%%%%%%%%%%%%%%%%%%%%%%%%%%%%%%%

\subsection{Impact of CMB constraints}

\begin{figure}[tb]
\centering
\hspace*{-0.6cm}\includegraphics[scale=0.88]{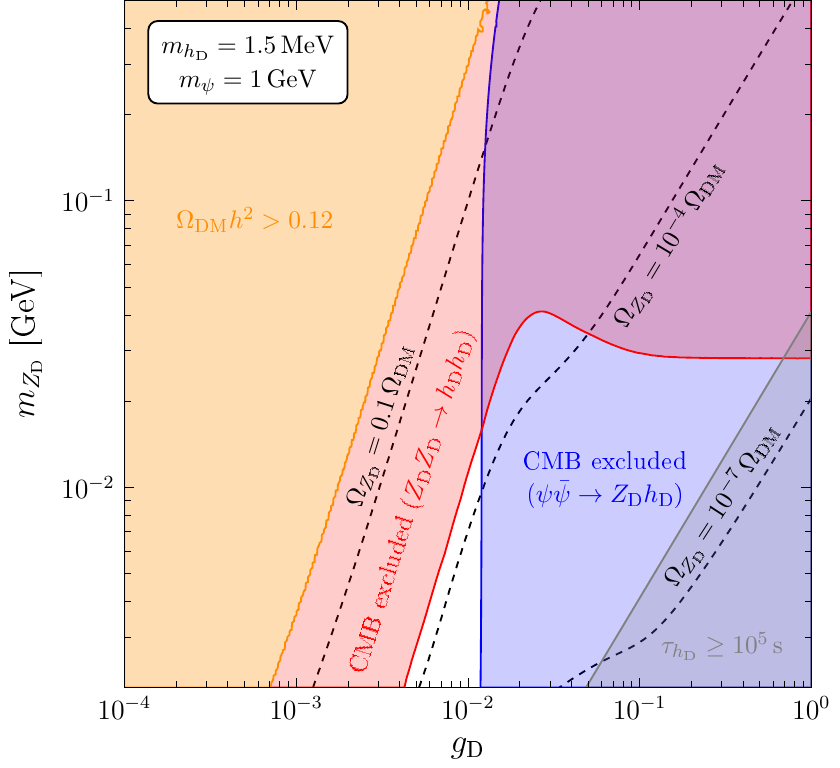}
\includegraphics[scale=0.88]{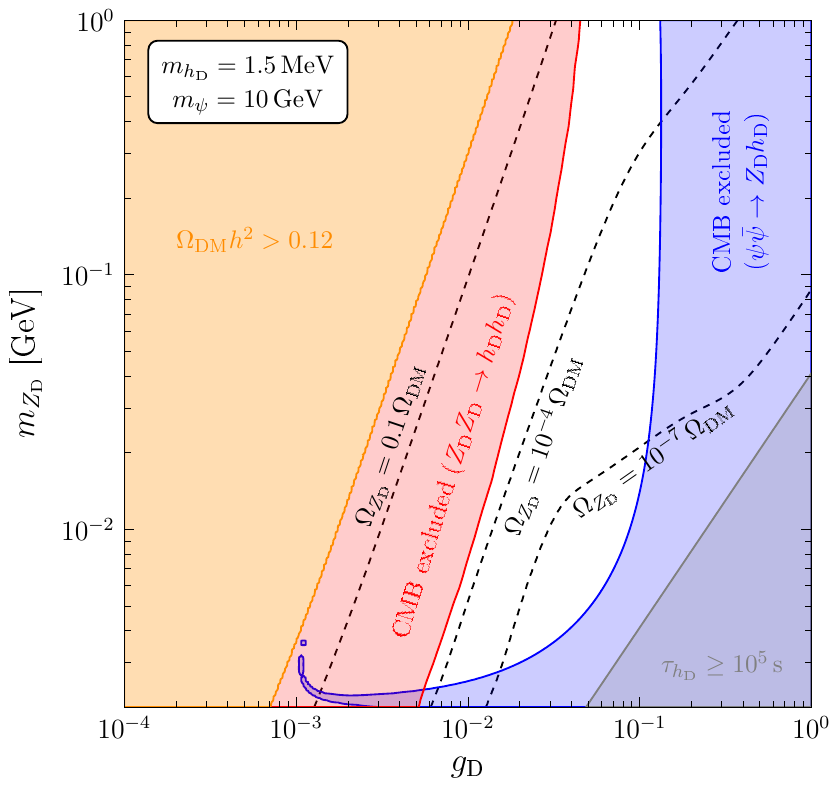}\\[0.45cm]
\hspace*{-0.6cm}\includegraphics[scale=0.88]{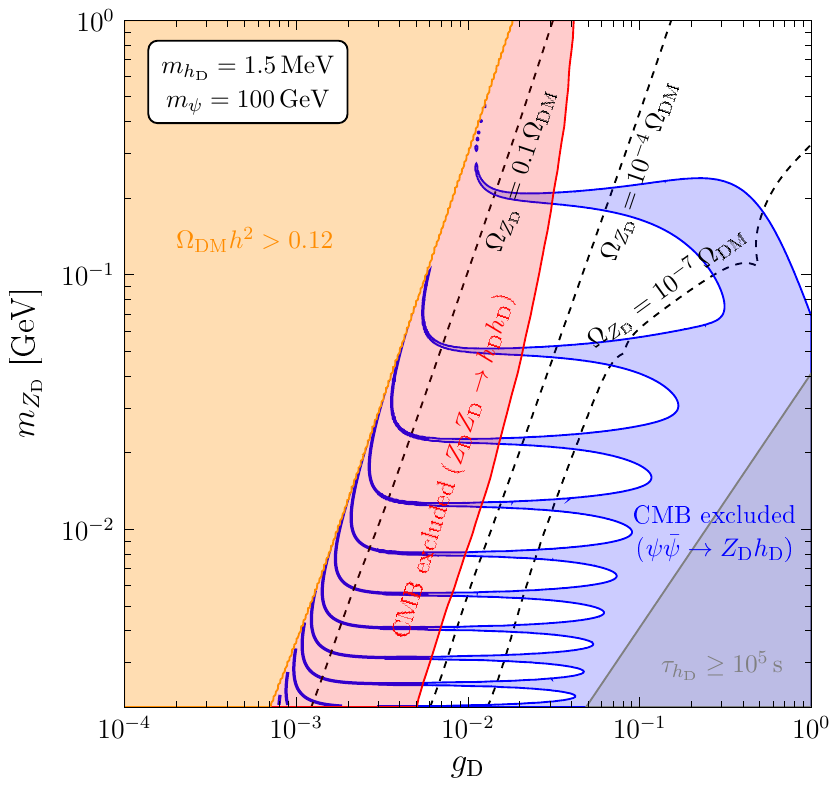}
\includegraphics[scale=0.88]{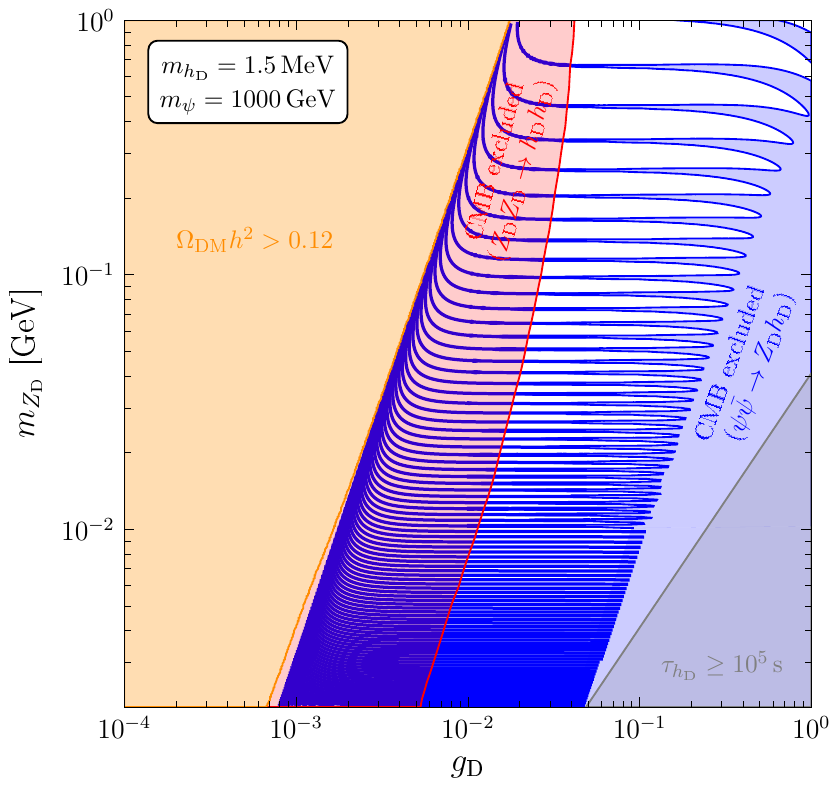}
\caption{Constraints in the $g_\text{D}$--$m_{Z_\text{D}}$ plane for $m_\psi = \unit[1]{GeV}$ (upper left panel), $m_\psi = \unit[10]{GeV}$ (upper right panel), $m_\psi = \unit[100]{GeV}$ (lower left panel), and $m_\psi = \unit[1000]{GeV}$ (lower right panel). In each case we fix $m_{h_\text{D}} = \unit[1.5]{MeV}$, and choose the coupling $g_\psi$ such that $\Omega_{\psi} h^2 + \Omega_{Z_\text{D}} h^2 \simeq 0.12$. In the orange shaded regions the DM density exceeds the observed value irrespective of the value of $g_\psi$. The regions of parameter space excluded by CMB constraints on late-time energy injection are given in blue for the process $\psi \bar{\psi} \to Z_\text{D} h_\text{D}$ and in red for $Z_\text{D} Z_\text{D} \to h_\text{D} h_\text{D}$. In the grey shaded areas, the lifetime of the dark Higgs $h_\text{D}$ exceeds $\unit[10^{5}]{s}$, leading to significant spectral distortions in the CMB. Note that the range of $m_{Z_\text{D}}$ shown for $m_\psi =  \unit[1]{GeV}$ (upper left plot) is smaller than in the rest of the panels.  \label{fig:gDmZDConstraints}}
\end{figure}

The CMB constraints on energy injection during recombination as discussed in section~\ref{sec:CMB} are illustrated in Fig.~\ref{fig:gDmZDConstraints}, where we show the parameter space spanned by the gauge coupling $g_\text{D}$ and the light DM mass $m_{Z_\text{D}}$ for different values of the mass of the heavy DM particle, $m_\psi =1$, 10, 100 and \unit[1000]{GeV}. Following the discussion in section~\ref{sec:BBN}, in order to evade constraints from spectral distortions of the CMB as well as from BBN as much as possible, we fix the mass of the dark Higgs boson to $m_{h_\text{D}} = \unit[1.5]{MeV}$, with the precise value being irrelevant to the CMB constraints on energy injection during recombination. Notice that with this choice one has $m_{h_\text{D}} < m_{Z_\text{D}}$ in all regions of parameter space shown in Fig.~\ref{fig:gDmZDConstraints}, as required for the annihilation channel $Z_\text{D} Z_\text{D} \to h_\text{D} h_\text{D}$ to be kinematically allowed. Lastly, the gauge coupling $g_\psi$ is fixed separately for each combination of $g_\text{D}$, $m_{Z_\text{D}}$ and $m_\psi$ by the requirement that $\psi$ and $Z_\text{D}$ together account for all of the observed DM, i.e.~$\Omega_{\text{DM}} h^2 \equiv \Omega_{\psi} h^2 + \Omega_{Z_\text{D}} h^2 \simeq 0.12$, following the discussion in  section~\ref{sec:DMAnnihilation}. The black dashed curves show contours of constant values of $\Omega_{Z_\text{D}} h^2/\Omega_{\text{DM}} h^2$, i.e.~the fraction of DM composed of $Z_\text{D}$. This fraction grows towards smaller values of $g_\text{D}$, until at some point the cross section for $Z_\text{D} Z_\text{D} \to h_\text{D} h_\text{D}$ [which scales as $g_\text{D}^4$, see eq.~\eqref{eq:ZDZDhdhD}] gets so small that irrespective of the choice of the gauge coupling $g_\psi$ controlling the relic density of $\psi$, the abundance of $Z_\text{D}$ alone overcloses the Universe. These regions of parameter space are shown as orange shaded in the different panels of Fig.~\ref{fig:gDmZDConstraints}.

In the blue shaded regions in Fig.~\ref{fig:gDmZDConstraints}, the energy injection from late-time annihilations $\psi \bar \psi \to Z_\text{D} h_\text{D}$ is excluded by CMB data, as defined in eq.~\eqref{eq:sigmav_NNX_bound}. Analogously, we show in red which parts of the parameter space are excluded by the CMB constraint on the annihilation process $Z_\text{D} Z_\text{D} \to h_\text{D} h_\text{D}$, c.f.~eq.~\eqref{eq:sigmav_ZDZDX_bound}. Finally, the grey-shaded regions are excluded on the basis of the lifetime of the dark Higgs boson ($\tau_{h_\text{D}} > \unit[10^{5}]{s}$), assuming $\lambda_{h \text{D}} = 4 \times 10^{-4}$ as discussed at the end of section~\ref{sec:BBN}.

For all values of $m_\psi$ shown in the different panels of Fig.~\ref{fig:gDmZDConstraints}, we find that the annihilations from the heavy and light DM particle constrain complementary regions in parameter space: the energy injection induced by the annihilation process $\psi \bar \psi \to Z_\text{D} h_\text{D}$ constrains  regions of parameter space with larger values of the gauge coupling $g_\text{D}$, while the bound derived from the annihilation of the lighter DM candidate $Z_\text{D}$ is most relevant for smaller $g_\text{D}$. This can be readily understood as follows: the annihilation cross section for $\psi \bar \psi \to Z_\text{D} Z_\text{D}$ (which does not lead to constraints from the CMB) scales with $g_\psi^4$, while $\psi \bar \psi \to Z_\text{D} h_\text{D}$ (which leads to the blue shaded exclusion regions in Fig.~\ref{fig:gDmZDConstraints}) is proportional to $g_\psi^2 g_\text{D}^2$. For sufficiently small values of $g_\text{D}$, the main annihilation channel of $\psi$ both during freeze-out and recombination is then given by the former process, and hence the CMB constraint from annihilations of $\psi$ becomes less and less important. On the other hand, for small values of $g_\text{D}$ the abundance of the lighter DM particle $Z_\text{D}$ is dominantly set by the annihilation process $Z_\text{D} Z_\text{D} \to h_\text{D} h_\text{D}$ (see appendix~\ref{app:relicdensity}), leading to $\Omega_{Z_\text{D}} h^2 \propto g_\text{D}^{-4}$. The corresponding bound from the CMB thus gets less important for larger values of $g_\text{D}$, as the suppression of the $Z_\text{D}$ abundance overcompensates the rise of the cross section towards larger values of the coupling: $(\Omega_{Z_\text{D}} h^2)^2 \times (\sigma v)_{Z_\text{D} Z_\text{D} \to h_\text{D} h_\text{D}} \propto g_\text{D}^{-8} \times g_\text{D}^4 = g_\text{D}^{-4}$. 

Interestingly, for all values of $m_\psi$ considered in Fig.~\ref{fig:gDmZDConstraints}, there remains a region of intermediate values of $g_\text{D}$ which is not constrained by either of the CMB constraints. Concretely, for $m_\psi = \unit[1]{GeV}$~(upper left panel), couplings in the interval $4 \times 10^{-3} \lesssim g_\text{D} \lesssim 10^{-2}$ are viable for $m_{Z_\text{D}} \sim \unit[2]{MeV}$, while all values of $g_\text{D}$ are excluded for $m_{Z_\text{D}} \gtrsim \unit[20]{MeV}$. Note that for this value of $m_\psi$, the region excluded by the process $Z_\text{D} Z_\text{D} \to h_\text{D} h_\text{D}$ becomes independent of the value of the coupling $g_\text{D}$ for the largest $Z_\text{D}$ masses shown. As discussed in more detail in appendix~\ref{app:relicdensity}, this is a result of additional annihilation channels significantly enhancing the abundance of $Z_\text{D}$ in this region of parameter space.

For larger values of $m_\psi$, we start to observe that the CMB constraint from the annihilation of $\psi$ reaches out to significantly smaller values of $g_\text{D}$ for specific values of $m_{Z_\text{D}}$. This is due to the resonant Sommerfeld enhancement of the annihilation process $\psi \bar{\psi} \to Z_\text{D} h_\text{D}$, occurring for $3 g_\psi^2 m_\psi = 2 \pi^3 n^2 m_{Z_\text{D}}$, with $n \in \mathbb{N}$~\cite{Cassel:2009wt} (see appendix~\ref{app:relicdensity} for more details). The larger the mass ratio $m_\psi/m_{Z_\text{D}}$, the closer these resonances are in parameter space, which becomes particularly visible in the lower right panel of Fig.~\ref{fig:gDmZDConstraints}, corresponding to $m_\psi = \unit[1]{TeV}$. In this case the values of $m_{Z_\text{D}}$ which are excluded or allowed by CMB constraints are extremely close to each other.\footnote{We note that for parameter points precisely on top of one of the Sommerfeld resonances, the calculation of the DM relic abundance might be affected by late-time annihilations not taken into account in our analysis~\cite{vandenAarssen:2012ag,Binder:2017lkj}.} We also note that when $g_\text{D}$ approaches the smallest value $g_\text{D}^\text{(min)}$ compatible with $\Omega_\text{DM} h^2 = 0.12$, the resonance peaks of the CMB constraint on $\psi \bar{\psi} \to Z_\text{D} h_\text{D}$ bend upwards. This is because in the limit $g_\text{D} \to g_\text{D}^\text{(min)}$, one has to lower the abundance of $\psi$ to ever smaller values in order to match the total DM abundance, implying increasingly larger values of $g_\psi$. \footnote{The required values of $g_\psi$ can become non-perturbative once $\Omega_{Z_\text{D}}$ is close to the observed DM relic density. While this may lead to a Landau pole below the Planck scale, this does not exclude further parts of the parameter space, as these regions are robustly excluded by the CMB constraints on $Z_\text{D}$ annihilation.} Thus, for fixed $m_{\psi}$, the resonance condition in this limit is satisfied for increasingly larger values of $m_{Z_\text{D}}$.

\begin{figure}[tb]
 \centering
 \hspace*{-0.6cm}\includegraphics[scale=0.88]{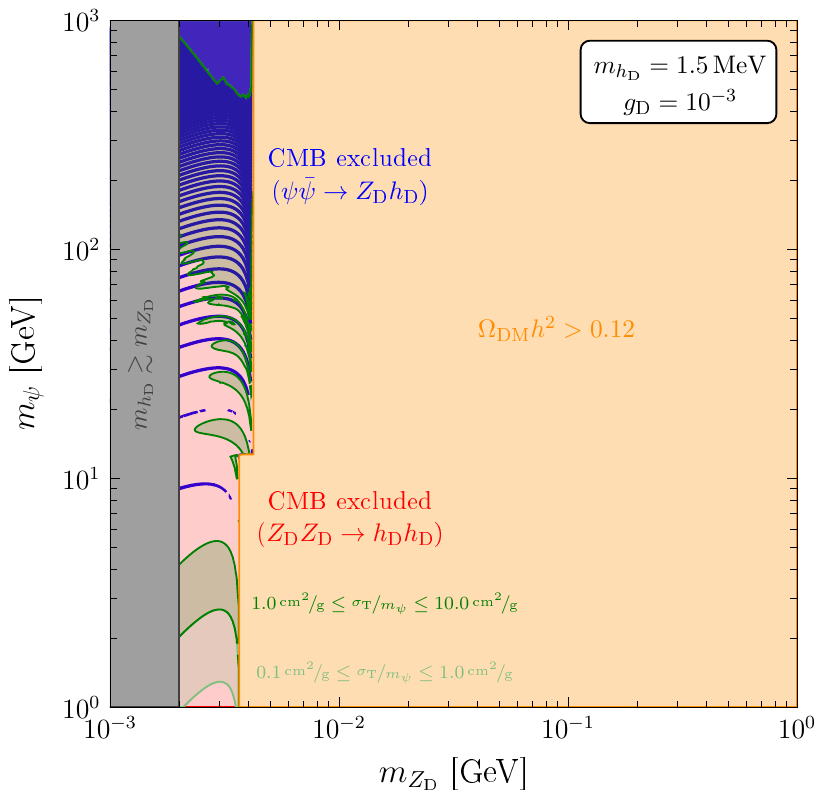}
 \includegraphics[scale=0.88]{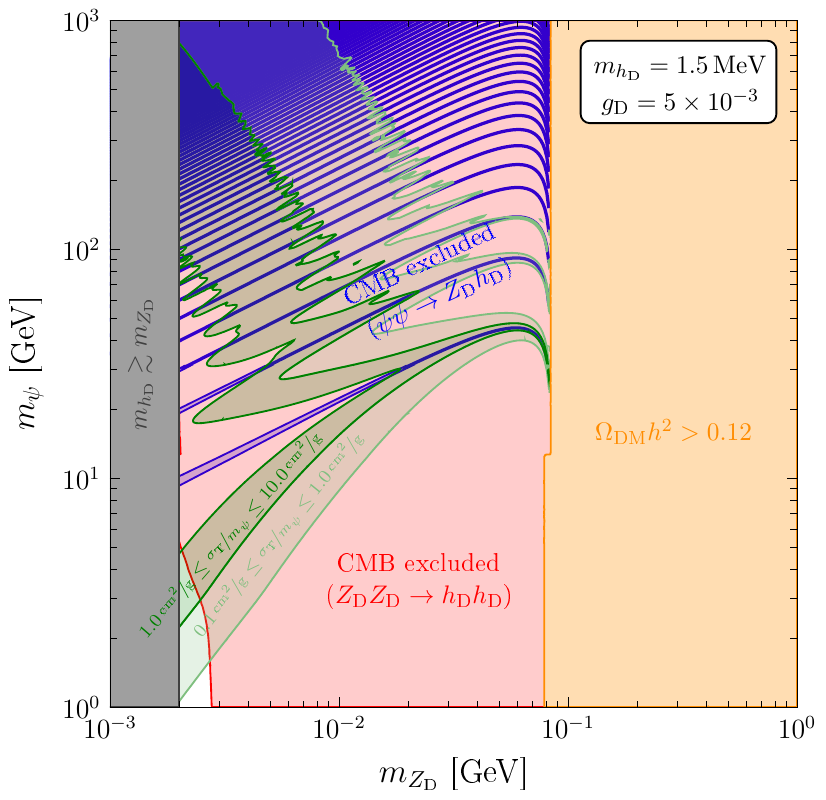}\\[0.45cm]
 \hspace*{-0.6cm}\includegraphics[scale=0.88]{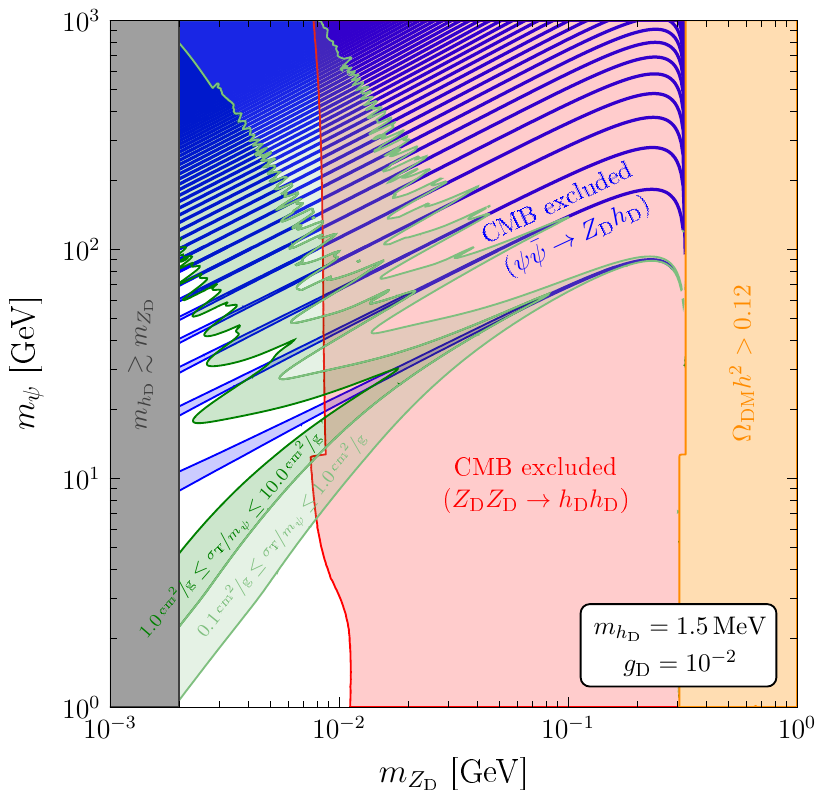}
 \includegraphics[scale=0.88]{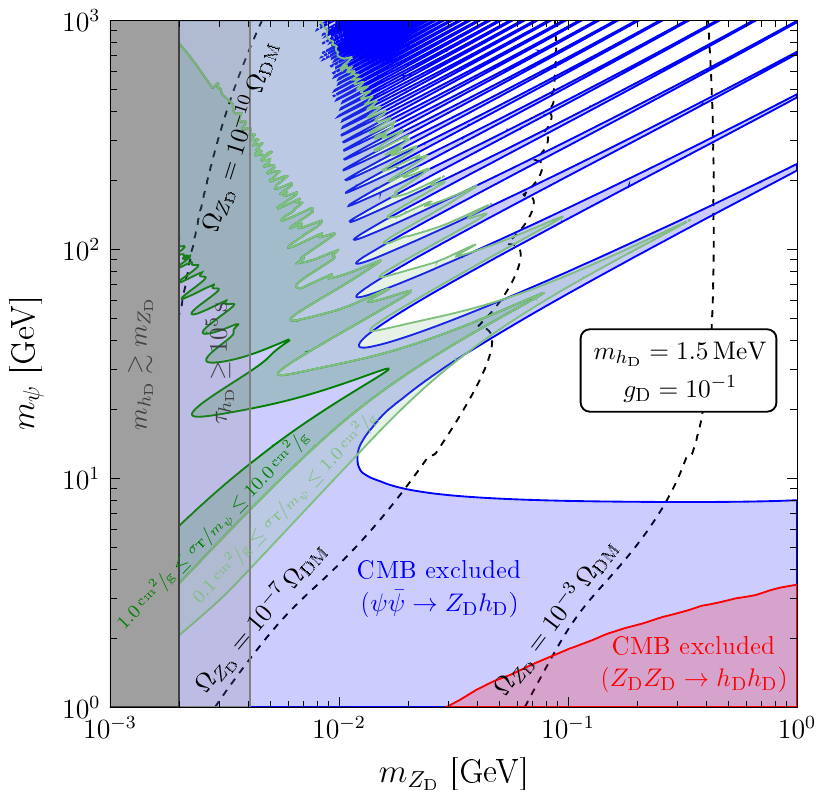}
 \caption{Constraints and regions of significant DM self-interaction cross section in the $m_{Z_\text{D}}$--$m_\psi$ plane for $g_\text{D} = 10^{-3}$ (upper left panel), $g_\text{D} = 5 \times 10^{-3}$ (upper right panel), $g_\text{D} = 10^{-2}$ (lower left panel) and $g_\text{D} = 10^{-1}$ (lower right panel). The coupling $g_\psi$ is fixed to reproduce the relic density where possible. As in Fig.~\ref{fig:gDmZDConstraints}, in the orange shaded regions one has $\Omega_\text{DM} h^2 > 0.12$, while the blue and red shaded regions indicate which parts of the parameter space are excluded by CMB constraints on energy injection from annihilation of $\psi$ and $Z_\text{D}$, respectively. In addition, we show in light (dark) green the combination of parameters leading to a self-interaction cross section of $\psi$ at the scale of dwarf galaxies in the range $0.1\, \text{cm}^2/\text{g} < \sigma_\text{T}/m_\psi < 1 \,\text{cm}^2/\text{g}$ ($1 \,\text{cm}^2/\text{g} < \sigma_\text{T}/m_\psi < 10\, \text{cm}^2/\text{g}$). The bound $\sigma_\text{T}/m_\psi \lesssim 1\, \text{cm}^2/\text{g}$ on the scale of galaxy clusters is satisfied in all of the parameter space shown in this figure.}
 \label{fig:mZDmNConstraints}
\end{figure}

\subsection{Viability of significant dark matter self-interactions}

Finally, in Fig.~\ref{fig:mZDmNConstraints} we present our results in the parameter space spanned by the masses $m_{Z_\text{D}}$ and $m_{\psi}$ of the two DM particles. From top left to bottom right, the four panels correspond to $g_\text{D} = 10^{-3}$, $5 \times 10^{-3}$, $10^{-2}$ and $10^{-1}$. Again, we fix $m_{h_\text{D}} = \unit[1.5]{MeV}$ and determine $g_\psi$ in each point of the parameter space by requiring the total DM density to be equal to the observed value.  As in Fig.~\ref{fig:gDmZDConstraints}, in the orange shaded regions the density of $Z_\text{D}$ is so large that $\Omega_{\text{DM}} h^2 > 0.12$ for all values of $g_\psi$. The blue and red shaded regions denote which combinations of parameters are excluded by the CMB constraint on energy injection from the annihilation of $\psi$ and $Z_\text{D}$, respectively.\footnote{The small discontinuity of the orange and red shaded region at $m_{\psi} \simeq \unit[12]{GeV}$ visible in some of the panels of Fig.~\ref{fig:mZDmNConstraints} is an artefact of our approximate treatment of the impact of the chemical decoupling of the visible and dark sector on the relic density of $Z_\text{D}$, c.f.~appendix~\ref{app:relicdensity}. A more precise treatment would lead to a smooth transition between the regions of different $m_\psi$, without affecting any of our conclusions.} In addition, we show in light and dark green the regions of parameter space leading to a self-interaction cross section of $\psi$ at the scale of dwarf galaxies in the range of $0.1\, \text{cm}^2/\text{g} < \sigma_\text{T}/m_\psi < 1 \,\text{cm}^2/\text{g}$ and $1 \,\text{cm}^2/\text{g} < \sigma_\text{T}/m_\psi < 10\, \text{cm}^2/\text{g}$, respectively. As outlined in section~\ref{sec:selfinteractions}, those values of $\sigma_\text{T}/m_\psi$ can potentially address the shortcomings of collisionless cold DM at small scales. On the other hand, the bound $\sigma_\text{T}/m_\psi \lesssim 1\,\text{cm}^2/\text{g}$ on the scale of galaxy clusters as discussed in section~\ref{sec:selfinteractions} is satisfied for the complete range of parameters shown in Fig.~\ref{fig:mZDmNConstraints}, and is thus not visible in the plots.

From the upper left panel of Fig.~\ref{fig:mZDmNConstraints} (corresponding to $g_\text{D}  = 10^{-3}$) it follows that for sufficiently small values of $g_\text{D}$, all of the parameter space leading to the interesting range of DM self-interaction cross sections at the scale of dwarf galaxies is excluded by CMB constraints on energy injection from the annihilation process $Z_\text{D} Z_\text{D} \to h_\text{D} h_\text{D}$. As already discussed above, this is a consequence of $Z_\text{D}$ contributing in a non-negligible way to the observed amount of DM for small values of $g_\text{D}$; concretely, we find $\Omega_{Z_\text{D}}/\Omega_\text{DM} \gtrsim 0.2$ for $g_\text{D}  = 10^{-3}$. On the other hand, the lower right panel of Fig.~\ref{fig:mZDmNConstraints} corresponding to  $g_\text{D}  = 0.1$ shows that if $g_\text{D}$ is sufficiently large, the bounds from the annihilation of $Z_\text{D}$ are irrelevant, but then most of the parameter space leading to the desired values of the self-interaction cross section $\sigma_\text{T}/m_\psi$ is excluded by CMB constraints arising from the annihilation process $\psi \bar \psi \to Z_\text{D} h_\text{D}$.

However, for intermediate values of the gauge coupling, such as $g_\text{D} = 10^{-2}$ shown in the lower left panel of Fig.~\ref{fig:mZDmNConstraints}, we indeed find regions in parameter space leading to $1 \,\text{cm}^2/\text{g} < \sigma_\text{T}/m_\psi < 10\, \text{cm}^2/\text{g}$ on the scale of dwarf galaxies and  $\sigma_\text{T}/m_\psi < 1 \,\text{cm}^2/\text{g}$ on the scale of galaxy clusters, while being consistent with the CMB bounds on the energy injection from the annihilation of $\psi$ and $Z_\text{D}$. Concretely, for $g_\text{D} = 10^{-2}$ this requires $m_{Z_\text{D}} \lesssim \unit[10]{MeV}$,\footnote{In these regions the values of $g_\psi$ are always within the perturbative regime, $g_\psi \in [0.01,0.5]$.} as well as a combination of $m_\psi$ and $m_{Z_\text{D}}$ sufficiently far away from one of the resonances corresponding to the narrow blue shaded regions in the plot. Let us remark again that even though for large values of $m_\psi$ the resonances are extremely dense in parameter space, the regions in between the resonance peaks are not excluded by CMB observations.

%%%%%%%%%%%%%%%%%%%%%%%%%%%%%%%%%%%%%%%%%%%%%%%%%%%%%%%%%%%%%%%%%%%%%%%%%%%%%%%
\section{Conclusions}
\label{sec:conclusions}
%%%%%%%%%%%%%%%%%%%%%%%%%%%%%%%%%%%%%%%%%%%%%%%%%%%%%%%%%%%%%%%%%%%%%%%%%%%%%%%

After years of theoretical and experimental efforts aiming at a better understanding of the astrophysical behaviour of DM at small scales, self-interacting DM remains one of the most compelling explanations for the apparent discrepancies found between observations and $N$-body simulations of collisionless cold DM. Realising the desired self-interaction cross section  $\sigma_\text{T}/m_\psi \simeq 1\,\text{cm}^2/\text{g}$ within a perturbative scenario of weak-scale DM requires the presence of a light mediator with a mass of $\unit[(0.1 - 100)]{MeV}$. However, two of the most basic incarnations of this general setup, a fermionic DM candidate coupled to an unstable scalar or vector mediator, are strongly disfavoured by the combination of data from direct detection experiments, CMB constraints on energy injection during recombination, as well as BBN constraints on late-time decaying particles.

In this article, we considered a scenario in which a \emph{stable vector mediator} $Z_\text{D}$ is responsible for the self-interactions of the fermionic DM particle $\psi$~\cite{Ma:2017ucp}. This immediately saves the model from CMB constraints on the annihilation process $\psi \bar \psi \to Z_\text{D} Z_\text{D}$. In order to suppress the cosmological abundance of the vector mediator to a level compatible with observations, we have introduced one more particle in the dark sector, a dark Higgs boson $h_\text{D}$ which is assumed to be lighter than $Z_\text{D}$. Besides being the natural by-product of the spontaneous breaking of a dark $U(1)$ gauge symmetry giving rise to the mass of the vector mediator, we have shown that the annihilation $Z_\text{D} Z_\text{D} \to h_\text{D} h_\text{D}$ can easily be efficient enough for $Z_\text{D}$ to only constitute a subdominant fraction of the observed DM. However, also this setup is subject to constraints from the CMB: the annihilation processes $\psi \bar \psi \to Z_\text{D} h_\text{D}$ as well as $Z_\text{D} Z_\text{D} \to h_\text{D} h_\text{D}$ together with the subsequent decay of the dark Higgs can lead to significant energy injection during recombination. Interestingly, we find that these two processes constrain complementary parts of the model parameter space, with the former being important only for sufficiently large values of the dark gauge coupling $g_\text{D}$ of the dark Higgs boson, and the latter for considerably smaller values. Most importantly, our results show that for a broad range of DM masses $m_\psi$ and $m_{Z_\text{D}}$, intermediate values of the gauge coupling $g_\text{D}$ ranging from $\sim 5 \times 10^{-3}$ to $\sim 10^{-1}$ are compatible with CMB constraints. 

Furthermore, we have discussed the constraints arising from the late-time decays of the thermally produced dark Higgs bosons. In order to evade the stringent bounds from CMB spectral distortions, the dark Higgs has to decay with a lifetime $\tau_{h_\text{D}} \lesssim \unit[10^5]{s}$, implying a mass $m_{h_\text{D}} > 2 m_e$. We have also discussed the possible impact of our scenario on the primordial abundances of light nuclei. For sufficiently small masses $m_{h_\text{D}} \lesssim \unit[4]{MeV}$, the decay products of the dark Higgs are not energetic enough to photo-disintegrate even the most weakly bound nucleus (deuterium), and consequently there are no constraints from BBN arising from late-time changes of the nuclear abundances. In addition, by setting the scalar coupling which is responsible for the mixing of the dark and SM Higgs boson to a value below $\simeq 4 \times 10^{-4}$, the dark and visible sector thermally decouple before the QCD phase transition, leading to a suppressed value of $\Delta N_\text{eff} \lesssim 0.27$ associated to the presence of $Z_\text{D}$ and $h_\text{D}$ in the thermal bath. Given all systematic uncertainties, this additional contribution to the energy density during BBN is still compatible with observations of primordial abundances.

Finally, we investigated whether the parts of parameter space which are compatible with all these constraints can lead to the range of desired values of the self-interaction cross section of DM at small scales. Indeed we find that for a gauge coupling $g_\text{D} \simeq 10^{-2}$, it is possible to obtain $1\,\text{cm}^2/\text{g} \lesssim \sigma_\text{T}/m_\psi \lesssim 10\,\text{cm}^2/\text{g}$ at the scale of dwarf galaxies, $\sigma_\text{T}/m_\psi \lesssim 1\,\text{cm}^2/\text{g}$ at the scale of galaxy clusters, while simultaneously being consistent with all CMB constraints on late-time energy injection as well as with BBN observations. In summary, our results thus show that if the scenario of DM interacting via an MeV-scale vector mediator is (minimally) extended by a dark Higgs boson breaking the dark gauge symmetry, it is indeed possible to restore the phenomenological viability of this setup in addressing the small-scale problems of the standard cold DM paradigm at small scales. Interestingly, the allowed range of parameters is already significantly narrowed down by current CMB and BBN observations, and could be further probed by future improvements of upper limits on the DM annihilation cross section at late times. In fact, the recent EDGES observation of an absorption feature in the 21 cm spectrum~\cite{Bowman:2018yin}, if confirmed, might already be able to supersede the CMB constraints on the DM annihilation cross section~\cite{DAmico:2018sxd,Liu:2018uzy,Cheung:2018vww}. Depending on the strength of the Sommerfeld enhancement at the relevant redshift $z \simeq 17$, the idea of DM self-interactions induced by the exchange of a light vector mediator as discussed in this article might thus be further probed in the near future.

\acknowledgments

We thank Camilo Garcia-Cely for useful discussions and Felix Kahlhoefer for valuable comments on the manuscript. This work is supported by the German Science Foundation (DFG) under the Collaborative Research Center (SFB) 676 Particles, Strings and the Early Universe as well as the ERC Starting Grant `NewAve' (638528).

\appendix

%%%%%%%%%%%%%%%%%%%%%%%%%%%%%%%%%%%%%%%%%%%%%%%%%%%%%%%%%%%%%%%%%%%%%%%%%%%%%%%
\section{Full Lagrangian}
\label{app:FullLagrangian}
%%%%%%%%%%%%%%%%%%%%%%%%%%%%%%%%%%%%%%%%%%%%%%%%%%%%%%%%%%%%%%%%%%%%%%%%%%%%%%%
 
In this appendix, we provide details of the Lagrangian~\eqref{eq:L} after the breaking of the dark and SM gauge symmetries by means of eq.~\eqref{eq:scalars_afterSB}. The portal term $\propto \lambda_{h \text{D}}$ appearing in eq.~\eqref{eq:scalarpotential_beforeSB} leads to a mixing of the scalar degrees of freedom $H_\text{D}$ and $H$. We define the mass eigenstates $h_\text{D}$ and $h$ via
\begin{align}
h_\text{D} &= - H \sin \theta + H_\text{D} \cos \theta \, ,\nonumber \\
h &= H \cos \theta + H_\text{D} \sin \theta \,,
\end{align}
with the mixing angle $\theta$ given by 
\begin{equation}
 \theta \simeq \lambda_{h\text{D}} v_\text{D} v_h/m_h^2 \,,
\end{equation}
assuming $\lambda_{h\text{D}} \ll 1$ and $m_{h_\text{D}} \ll m_h$. Trading the parameters $\mu_\text{D}$ and $\mu_h$ of the Higgs potential for the physical masses $m_{h_\text{D}}$ and $m_h$, and replacing $v_\text{D}$ by $m_{Z_\text{D}}/g_\text{D}$, the scalar potential including only the leading terms in an expansion in $\lambda_{h\text{D}}$ reads
\begin{align}
V_{\text{broken}} (h, h_\text{D}) &\simeq \frac12 m_h^2 h^2 + \frac{m_h^2}{2 v_h} h^3 + \frac{m_h^2}{8 v_h^2} h^4 + \frac12 m_{h_\text{D}}^2 h_\text{D}^2 + \frac{g_\text{D} m_{h_\text{D}}^2}{2 m_{Z_\text{D}}} h_\text{D}^3 + \frac{g_\text{D}^2 m_{h_\text{D}}^2}{8 m_{Z_\text{D}}^2} h_\text{D}^4 \nonumber \\
&\quad +\frac12 \lambda_{h\text{D}} v_h h h_\text{D}^2 - \frac{\lambda_{h\text{D}} m_{Z_\text{D}}}{g_\text{D}} h^2 h_\text{D} + \frac14  \lambda_{h\text{D}} h^2 h_\text{D}^2 \nonumber \\
&\quad + \frac{\lambda_{h\text{D}} g_\text{D} v_h m_{h_\text{D}}^2}{2 m_{Z_\text{D}} m_h^2} h h_\text{D}^3 - \frac{\lambda_{h\text{D}} m_{Z_\text{D}}}{2 g_\text{D} v_h} h^3 h_\text{D} \,.
\end{align}

The full Lagrangian after symmetry breaking is then finally given by
\begin{align}
\mathcal{L} &\simeq \mathcal{L}_{\widetilde{\text{SM}}} \big|_{H \rightarrow h - \theta h_\text{D}} - \frac{1}{4} F^{\mu\nu}_\text{D} F_{\mu\nu}^D + \frac12 m_{Z_\text{D}}^2 Z_\text{D}^\mu Z_{D\mu} +  i \bar{\psi} \gamma_\mu \partial^\mu \psi +  g_\psi \bar{\psi} \gamma_\mu Z_\text{D}^\mu \psi - m_\psi \bar{\psi} \psi \nonumber \\
&\quad\quad + g_\text{D} m_{Z_\text{D}} (h_\text{D} + \theta h) Z_\text{D}^\mu Z_{D\mu} + \frac12 g_\text{D}^2 (h_\text{D} + \theta h)^2 Z_\text{D}^\mu Z_{D\mu} \nonumber \\
&\quad\quad+\frac12 (\partial^\mu h_\text{D}) (\partial_\mu h_\text{D}) - V_{\text{broken}} (h, h_\text{D}) \,.
\label{eq:fullL_afterSB}
\end{align}
Notice that here we neglect the modifications proportional to $\theta^2$ of couplings of $h$ to SM fields. The couplings of $h_\text{D}$ to the SM gauge bosons $Z$ and $W$, as well as to the SM fermions $f$ are given by
\begin{multline}
\mathcal{L}_{\widetilde{\text{SM}}} \big|_{H \rightarrow h - \theta h_\text{D}} \supset \theta \left( \sum_f  \frac{m_f}{v_h} \bar f f h_\text{D} \right)  \\
+  \frac{\theta m_Z^2}{2 v_h^2} \times  \left( -2  v_h h_\text{D} - 2 h h_\text{D} + \theta h_\text{D}^2\right) \left( Z_\mu Z^\mu  + 2 \cos^2 \theta_W  W_\mu^+ W^{\mu-}\right),
\end{multline}
with $\theta_W$ the Weinberg angle. 

%%%%%%%%%%%%%%%%%%%%%%%%%%%%%%%%%%%%%%%%%%%%%%%%%%%%%%%%%%%%%%%%%%%%%%%%%%%%%%%
\section{Relic density calculation}
\label{app:relicdensity}
%%%%%%%%%%%%%%%%%%%%%%%%%%%%%%%%%%%%%%%%%%%%%%%%%%%%%%%%%%%%%%%%%%%%%%%%%%%%%%%

In this appendix we describe in detail our method for calculating the relic abundances of the two DM particles $\psi$ and $Z_\text{D}$ for a given point in parameter space. In particular, we discuss the treatment of Sommerfeld enhancement during freeze-out, the importance of DM conversion and semi-annihilation processes, as well as the chemical decoupling of the dark and visible sector during or after DM freeze-out.

We implemented the Lagrangian of the model with \texttt{FeynRules v2.3.24}~\cite{Alloul:2013bka} and generated \texttt{CalcHEP}~\cite{Belyaev:2012qa} model files to be imported into \texttt{MicrOMEGAs v4.3.5}~\cite{Belanger:2006is,Belanger:2014vza}. However, we find that due to the large mass hierarchy between the initial and final state particles, e.g.~in the annihilation process $\psi \bar \psi \rightarrow Z_\text{D} Z_\text{D}$, the calculation of the annihilation cross sections using \texttt{CalcHEP} is facing numerical problems related to the polarisation sums over the light massive vector particles.\footnote{See appendix C.2 of Ref.~\cite{Belyaev:2012qa} for a detailed discussion of this point.} We therefore compute all relevant annihilation cross sections analytically and pass them to \texttt{MicrOMEGAs} for further use in the numerical solution of the Boltzmann equations. In doing so, we also take into account the Sommerfeld enhancement in the annihilation processes $\psi \bar \psi \rightarrow Z_\text{D} Z_\text{D}$ and $\psi \bar \psi \rightarrow Z_\text{D} h_\text{D}$, arising from the multiple exchange of the light vector boson $Z_\text{D}$ in the initial state~\cite{sommerfeld}. In practice, we compute the $s$- and $p$-wave contributions to the corresponding annihilation cross sections  at tree level, and multiply them with enhancement factors $S_s$ and $S_p$, respectively. Following~\cite{Cassel:2009wt,Iengo:2009ni,Slatyer:2009vg}, we approximate the Yukawa potential generated by the exchange of $Z_\text{D}$ by a Hulth\'en potential, leading to
\begin{align}
 S_s &= \frac{\pi}{a} \frac{\sinh (2 \pi  a  c)}{\cosh (2 \pi  a  c) - \cos (2 \pi \sqrt{c - a^2 c^2})} \; , \label{eq:Ss}\\
 S_p &= \frac{(c-1)^2 + 4 \, a^2 c^2}{1 + 4 \, a^2 c^2} \times S_s \; ,
\end{align}
where $a = 2 \pi v/g_\psi^2$ and  $c = 3 g_\psi^2 m_\psi/(2 \pi^3 m_{Z_\text{D}})$.

The Boltzmann equations for the number densities $n_\psi$ and $n_{Z_\text{D}}$ are then given by
\begin{align}
\left( \frac{\text{d}n_\psi}{\text{d}t} + 3 H n_\psi \right)\bigg|_{T \gg m_{Z_\text{D}}} \simeq &- \left( \langle \sigma v\rangle_{\psi \bar \psi \to Z_\text{D} Z_\text{D}} + \langle \sigma v\rangle_{\psi \bar \psi \to Z_\text{D} h_\text{D}} \right) \left( n_\psi^2 - \overline{n}_\psi^2\right)\,, \label{eq:Boltzmann_npsi}\\[0.3cm]
\left( \frac{\text{d}n_{Z_\text{D}}}{\text{d}t} + 3 H n_{Z_\text{D}} \right) \bigg|_{T \lesssim m_{Z_\text{D}} \ll m_\psi} \simeq &- \langle \sigma v\rangle_{Z_\text{D} Z_\text{D} \to h_\text{D} h_\text{D}} \left( n_{Z_\text{D}}^2 - \overline{n}_{Z_\text{D}}^2\right) \nonumber\\
&+ \left( \langle \sigma v\rangle_{\psi \bar \psi \to Z_\text{D} Z_\text{D}} + \frac12 \langle \sigma v\rangle_{\psi \bar \psi \to Z_\text{D} h_\text{D}} \right) n_{\psi}^2 \nonumber \\
&-\frac12 \langle \sigma v\rangle_{\psi Z_\text{D} \to \psi h_\text{D}} \left(n_{Z_\text{D}} - \overline{n}_{Z_\text{D}}\right) n_\psi \,, \label{eq:Boltzmann_nZD}
\end{align}
with $\overline{n}_\psi$ and $\overline{n}_{Z_\text{D}}$ denoting number densities in equilibrium, and $H$ being the Hubble rate. For the sake of the following discussion, in  these expressions (but not in our numerical calculation\footnote{\texttt{MicrOMEGAs} solves the full Boltzmann equations in the temperature interval \texttt{[Tstart,Tend]}. In order to make sure that the freeze-out of $Z_\text{D}$ occurs within this range of temperatures even for the smallest values of $m_{Z_\text{D}}$ considered in this work, we lower \texttt{Tend} from the default value $ \unit[10^{-3}]{GeV}$ to  $ \unit[10^{-6}]{GeV}$.}) we have set $\overline{n}_{Z_\text{D}} \simeq n_{Z_{\text{D}}}$ during freeze-out of $\psi$, as well as $\overline{n}_\psi \simeq 0$ during the freeze-out process of $Z_{\text{D}}$. Under these assumptions, which are fulfilled to good accuracy as long as $m_{Z_\text{D}} \ll m_\psi$, the Boltzmann equation for $n_\psi$ takes the same form as in the standard scenario of a single DM particle and can be solved independently of the evolution of $n_{Z_\text{D}}$. 

On the other hand, the final abundance of the lighter DM particle $Z_\text{D}$ can be significantly affected by the additional terms in eq.~\eqref{eq:Boltzmann_nZD} involving the heavy DM particle $\psi$ (see also~\cite{Ahmed:2017dbb}). For the case of the annihilation processes $\psi \bar \psi \to Z_{\text{D}} Z_{\text{D}} $ and $\psi \bar \psi \to Z_{\text{D}} h_{\text{D}}$, this can be qualitatively understood by considering the ratio of the second and first term in the Boltzmann equation, evaluated at the temperature $T_f$ where the annihilation process $Z_{\text{D}} Z_{\text{D}} \to h_{\text{D}} h_{\text{D}}$ falls out of equilibrium:
\begin{align}
\kappa_{\psi \bar \psi} &\equiv \frac{\left( \langle \sigma v\rangle_{\psi \bar \psi \to Z_\text{D} Z_\text{D}} + \frac12 \langle \sigma v\rangle_{\psi \bar \psi \to Z_\text{D} h_\text{D}} \right) \cdot n_{\psi}^2(T_f)}{\langle \sigma v\rangle_{Z_\text{D} Z_\text{D} \to h_\text{D} h_\text{D}} \cdot n_{Z_\text{D}}^2(T_f)} \nonumber \\
&= \frac{\left( \langle \sigma v\rangle_{\psi \bar \psi \to Z_\text{D} Z_\text{D}} + \frac12 \langle \sigma v\rangle_{\psi \bar \psi \to Z_\text{D} h_\text{D}} \right) \cdot Y_{\psi}^2(T_f)}{\langle \sigma v\rangle_{Z_\text{D} Z_\text{D} \to h_\text{D} h_\text{D}} \cdot Y_{Z_\text{D}}^2(T_f)} \,,
\label{eq:kappa}
\end{align}
where in the second line we replaced the number densities $n$ by the yields $Y = n/s$. If $\kappa_{\psi \bar \psi} \gtrsim 1$, the standard calculation for the freeze-out of $Z_\text{D}$ only taking into account the annihilation process $Z_\text{D} Z_\text{D} \to h_\text{D} h_\text{D}$ fails, as the residual annihilations of $\psi$ contribute significantly to the production of $Z_\text{D}$ around the freeze-out temperature $T_f$. In regions of parameter space where $\psi$ is the dominant component of DM, the numerator of eq.~\eqref{eq:kappa} can be estimated by setting the total annihilation cross section of $\psi$ to $\langle \sigma v \rangle_{\text{thermal}} \simeq 4.4 \times 10^{-26} \, \text{cm}^3\,\text{s}^{-1}$, and the yield $Y_\psi$ to the value corresponding to $\Omega_\psi h^2 \simeq 0.12$. Furthermore, an approximate expression for $Y_{Z_\text{D}}(T_f)$ can be obtained from the semi-analytical solution to the standard one-particle Boltzmann equation~\cite{Kolb:1990vq}:
\begin{align}
Y_{Z_\text{D}}(T_f) &\simeq  \frac{3.79}{\sqrt{g_\star} M_\text{P} m_{Z_\text{D}} \sigma_0} \log \left( \frac{0.11}{\sqrt{g_\star}} M_\text{P} m_{Z_\text{D}} \sigma_0 \right) \,,
\end{align}
where $g_\star \simeq 10.75$ denotes the SM degrees of freedom at $T_f$, $M_\text{P}$ is the Planck mass, and $\sigma_0 \equiv (\sigma v)^{v \rightarrow 0}_{Z_\text{D} Z_\text{D} \to h_\text{D} h_\text{D}}$. Finally, after inserting the analytical expression for $\sigma_0$ given in eq.~\eqref{eq:ZDZDhdhD} we obtain
\begin{align}
 \kappa_{\psi \bar \psi} \simeq 0.029 \cdot \left( \frac{g_\text{D}}{10^{-2}} \right)^4 \left( \frac{m_\psi}{\text{GeV}} \right)^{-2} \cdot \left( 1 + 0.16 \log \left[  \frac{g_\text{D}}{10^{-2}} \right] - 0.040 \log \left[ \frac{m_{Z_\text{D}}}{\text{MeV}} \right] \right)^{-2}\,,
 \label{eq:kappa_eval}
\end{align}
assuming $m_{h_\text{D}} \ll m_{Z_\text{D}}$. Clearly, for sufficiently large $g_\text{D}$ and small $m_\psi$ one has $\kappa_{\psi \bar \psi} \gtrsim 1$, indicating that the annihilation processes of the heavy DM particle $\psi$ should indeed be taken into account in the calculation of the relic abundance of $Z_\text{D}$.

\begin{figure}[tb]
\centering
\includegraphics[scale=1.2]{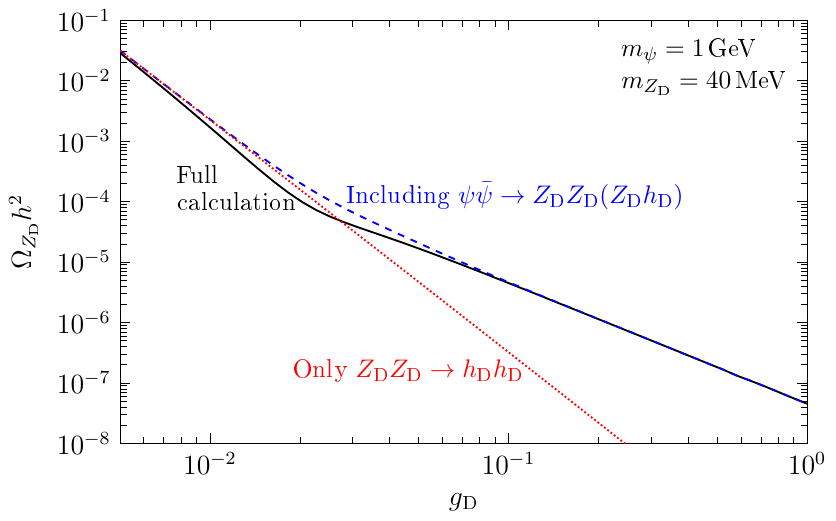}
\caption{Relic abundance of $Z_\text{D}$ as a function of $g_\text{D}$, assuming $m_\psi = \unit[1]{GeV}$ and $m_{Z_\text{D}} = \unit[40]{MeV}$. The red dotted curve only takes into account the annihilation $Z_\text{D} Z_\text{D} \to h_\text{D} h_\text{D}$, the blue dashed curve in addition the self-annihilation of $\psi$, and the black solid curve corresponds to the full calculation including all terms of eq.~\eqref{eq:Boltzmann_nZD}.}
\label{fig:OmegaPlot}
\end{figure}

These simple analytical considerations are confirmed using our full numerical approach of solving the Boltzmann equation via \texttt{MicrOMEGAs}.  In Fig.~\ref{fig:OmegaPlot} we show the relic abundance of $Z_\text{D}$ as a function of the coupling $g_\text{D}$, fixing for concreteness $m_\psi = \unit[1]{GeV}$ and $m_{Z_\text{D}} = \unit[40]{MeV}$. The red dotted curve corresponds to a calculation where only the annihilation process $Z_\text{D} Z_\text{D} \to h_\text{D} h_\text{D}$ is taken into account; as expected, the corresponding abundance scales as $\Omega_{Z_\text{D}} h^2 \propto 1/\langle \sigma v\rangle_{Z_\text{D} Z_\text{D} \to h_\text{D} h_\text{D}} \propto g_\text{D}^{-4}$. On the other hand, the blue dashed curve shows the abundance obtained by additionally including the terms in the Boltzmann equation accounting for the self-annihilation of $\psi$. The two calculations deviate significantly once $g_\text{D} \gtrsim 2\times 10^{-2}$, well compatible with the simple estimate based on eq.~\eqref{eq:kappa_eval}. Lastly, the solid black curve furthermore takes into account the conversion process  $\psi Z_\text{D} \to \psi h_\text{D}$, which impacts the calculation mainly for intermediate values of $g_\text{D}$.

Finally, we take into account the impact of the thermal decoupling of the visible and dark sector on the abundances of $\psi$ and $Z_\text{D}$. As explained in section~\ref{sec:BBN}, in order to evade the bounds from BBN and CMB spectral distortions as much as possible, we fix $\lambda_{h \text{D}} \simeq 4 \times 10^{-4}$ in our analysis. Then, as shown in Fig.~\ref{fig:hDhD_rates}, the dark and visible sector decouple at $T_\text{dec} \simeq \unit[500]{MeV}$. Assuming separate entropy conservation in both sectors for $T < T_\text{dec}$, the dark sector temperature $T_\text{D}$ as a function of the photon temperature $T$ evolves according to
\begin{equation}
 \xi(T) \equiv \frac{T_\text{D}(T)}{T}=\left(\frac{g_{\ast S} (T)}{g_{\ast S} (T_\text{dec})} \, \frac{g_{\ast S}^\text{D} (T_\text{dec})}{g_{\ast S}^\text{D} (T_\text{D})}\right)^\frac{1}{3} \,,
\end{equation}
where $g_{\ast S}(T)$ and $g_{\ast S}^\text{D}(T_\text{D})$ denote the entropy degrees of freedom in the visible and dark sector at a given temperature. For the range of particle masses considered in our analysis, $Z_\text{D}$ always freezes out after the decoupling of the two sectors, and so does $\psi$ for $m_\psi \lesssim \unit[12]{GeV}$. Following~\cite{Feng:2008mu}, we take this into account by applying separate correction factors $\xi(T_\text{fo})$ to the relic abundances of $\psi$ and $Z_\text{D}$ obtained from a calculation assuming equal temperatures in both sectors, where $T_\text{fo}$ is the freeze-out temperature of $\psi$ or $Z_\text{D}$, respectively. Note that we implicitly assume $h_\text{D}$ to be a relativistic degree of freedom to ensure $g_{\ast S}^\text{D} (T_\text{D}) > 0$; possible corrections to the abundance of $Z_\text{D}$ in situations where all particles in the dark sector have become non-relativistic during freeze-out (see e.g.~\cite{Pappadopulo:2016pkp}) are left for future work.

\bibliographystyle{JHEP_improved}
\bibliography{SIDM_StableMediator}

\end{document}